\documentclass[a4paper,fleqn]{cas-dc}

\usepackage[authoryear]{natbib}
\usepackage{xcolor}
\usepackage{placeins}

\def\tsc#1{\csdef{#1}{\textsc{\lowercase{#1}}\xspace}}
\tsc{WGM}
\tsc{QE}
\tsc{EP}
\tsc{PMS}
\tsc{BEC}
\tsc{DE}
\defcitealias{wiersma_effect_2009}{W09}
\newcommand{\wie}{\citetalias{wiersma_effect_2009}}

\begin{document}

\shorttitle{Net Cooling Tables for Cosmological Simulations}
\shortauthors{Di Federico et al.}

\title [mode = title]{Net Cooling Tables for Cosmological Simulations under Multiple UV/X-ray Background Models}

\author[USM]{Laura   Di Federico}

\credit{Conceptualization of this study, Methodology, Software, Writing}

\author[USM, MPA]{Klaus Dolag}

\credit{}

\author[RPC, USM, MPA]{Ildar Khabibullin}

\credit{}

\author[USM]{Luca Sala}

\credit{}

\author[USM]{Geray S. Karademir}

\credit{}

\author[UNITS,INAFTS,ICSC,IFPU]{Milena Valentini} 

\credit{}

%\address[1]{, Street 129, 1043 NX Amsterdam, The Netherlands}
\affiliation[USM]{organization={University Observatory, Faculty of Physics, Ludwig-Maximilians-Universität München},
            addressline={Scheinerstr. 1},
            city={Munich},
            postcode={81679},
            country={Germany}}

\affiliation[MPA]{organization={Max-Planck-Institute for Astrophysics},
            addressline={Karl-Schwarzschild-Straße 1},
            city={Garching},
            postcode={85741},
            country={Germany}} 
            
\affiliation[RPC]{organization={Rudolf Peierls Centre for Theoretical Physics, Department of Physics, University of Oxford},
                addressline={Parks Rd},
                city={Oxford}, 
                postcode={OX1 3PU},
                country={United Kingdom}}

\affiliation[UNITS]{organization={Dipartimento di Fisica, Università di Trieste}, 
            addressline={via Valerio 2},
            city={Trieste},
            postcode={I-34127}, 
            country={Italy}}

\affiliation[INAFTS]{organization={INAF -- Osservatorio Astronomico di Trieste}, 
            addressline={via Tiepolo 11},
            city={Trieste},
            postcode={I-34143}, 
            country={Italy}}
                       
\affiliation[ICSC]{organization={ICSC - Italian Research Center on High Performance Computing, Big Data and Quantum Computing},
            addressline={via Magnanelli 2},
            city={Casalecchio di Reno},
            postcode={40033},
            country={Italy}}

\affiliation[IFPU]{organization={IFPU, Institute for Fundamental Physics of the Universe},
            addressline={via Beirut 2},
            city={Trieste},
            postcode={34014},
            country={Italy}}      

% ABSTRACT
\begin{abstract}
We present three sets of net cooling tables for cosmological hydrodynamical simulations, computed with CLOUDY 23.01 for optically thin gas in photoionization equilibrium with three different UV/X-ray background models. Following the element-by-element approach, and in a format compatible with existing tables, we tabulate cooling, heating, and net rates separately for hydrogen and helium over an extended range of helium mass fractions and for 13 heavy elements, together with hydrogen species fractions, electron abundances, and mean particle masses. For gas of Solar composition, the choice of UV/X-ray background has little impact on the net cooling rates at low redshift, but the results increasingly diverge towards high redshift, particularly in diffuse gas near thermal equilibrium. Compared with widely used previous tables, cooling times in this regime differ by up to an order of magnitude at $z \gtrsim 4.6$, and the calculations disagree on whether the gas undergoes net cooling or heating over nearly 10\% of the diffuse density-temperature plane at $z \approx 8$, corresponding to the intergalactic medium. These trends persist for gas composed exclusively of hydrogen and helium. The tables provide a consistent framework for assessing the impact of the radiation background on gas cooling in cosmological simulations.
\end{abstract}

% KEYWORDS

\maketitle

% INTRODUCTION

\section{Introduction}
\label{sec:introduction}

Radiative cooling is a fundamental process in many different contexts in astrophysics, particularly in the formation and evolution of cosmic structures. As gas falls into gravitational potential wells, it is compressed and shock-heated \citep{Birnboim2003, Kerevs2005}, and its subsequent evolution depends on how efficiently this thermal energy can be radiated away. The cooling efficiency is determined by the chemical composition and ionization state of the gas, and defines the characteristic cooling time. The ability of the gas to condense then depends on the comparison between this timescale and the relevant dynamical timescale of the halo \citep{rees_cooling_1977, Silk1977, white_core_1978}: if cooling is sufficiently rapid, the gas can lose pressure support, condense, and ultimately form stars.

Once stars form, stellar winds and supernovae enrich the surrounding gas with heavy elements \citep{CenOstriker, Tornatore2007, WiersmaBIS, Nomoto2013}. This enrichment can strongly increase radiative cooling by providing many additional electronic transitions. At the same time, radiation from stars photoionizes and heats the gas, reducing its ability to cool \citep{wiersma_effect_2009}. As structure formation proceeds, active galactic nuclei (AGN) provide an additional source of ionizing radiation. This radiation propagates through the inhomogeneous intergalactic medium (IGM), where it is absorbed by discrete absorbers spanning a wide range of column densities \citep{HM96, haardt_radiative_2012}. These absorbers are not only sinks but also sources of ionizing photons, since a significant fraction of the absorbed radiation is re-emitted through recombinations \citep{HM96, 2009ApJ...703.1416F}. These sources and sinks together shape the radiation field to which the gas is exposed.

Photoionization is not the only process that can heat the gas. Compton scattering between photons and free electrons transfers energy from the photons to the gas when the photon energies exceed the thermal energies of the electrons, as in the presence of X-ray emission from AGN \citep{sazonov_quasars_2004}. In the opposite regime, hot electrons transfer energy to lower-energy photons through inverse Compton scattering, as in the scattering of cosmic microwave background (CMB) photons by hot intracluster gas \citep{Sunyaev1972}. Cosmic rays, relativistic particles produced by supernova remnants and AGN, can also heat the gas by ionizing atoms and dissociating molecules. This process is especially important in molecular clouds \citep{Padovani2009}, which are self-shielded from UV photons.

A realistic description of the thermal evolution of the gas in a cosmological framework therefore requires considering the net balance between radiative cooling and heating. This balance depends on the physical properties and chemical composition of the gas, as well as on the radiation field to which it is exposed. In principle, calculating this balance requires following the relevant atomic processes for each ionic species and accurately treating the radiation field. This radiation field itself depends on the luminosities, spectra, distances, and geometrical distribution of the sources and on the sinks. 

Performing these calculations self-consistently at every time step is computationally too expensive for cosmological simulations. Cooling and heating rates are therefore commonly precomputed using simplified assumptions and stored in tables. These are then interpolated for each gas particle at each time step in order to update the internal energy. This update can be performed with explicit \citep[e.g.\ \textsc{grackle};][]{Grackle17} or implicit \citep[e.g.][]{Hernquist89, gadget2001} integration schemes, or with an exact integration along the tabulated cooling curves \citep{Townsend09, zhu_gas_2017}. In the latter case, the accuracy of the thermal evolution is limited by the resolution of the tables rather than by the time step. 

The initial calculations of the cooling rate assumed that the gas was optically thin and in collisional ionization equilibrium (CIE). In the CIE regime, the ionization state of the gas is determined by the balance between collisional ionization and recombination and therefore depends on the temperature and chemical composition. In a primordial hydrogen (H) and helium (He) gas, cooling arises mainly from collisional excitation and ionization, recombination, and free-free emission. The inclusion of heavy elements adds metal-line emission, which can strongly increase cooling at$10^{4}\lesssim T\lesssim10^{7}$~K \citep[e.g.][]{cox_ionization_1969, raymond_radiative_1976, boehringer_metallicity_1989, sutherland_cooling_1993, gnat_time-dependent_2007}. At higher temperatures, the gas becomes almost fully ionized and free-free emission becomes increasingly important.

CIE tables provide a useful description of collisionally ionized gas, but they do not account for the effect of an external ionizing radiation field on the ionization state of the gas. Cosmological calculations commonly describe this radiation using a spatially uniform, redshift-dependent ultraviolet and X-ray background (UVB). This avoids the need to follow the radiation from individual sources. Cooling and heating rates are tabulated as functions of density, temperature, and chemical composition. The ionization state in this regime assumes photoionization equilibrium (PIE), in which the balance is determined by photoionization, collisional ionization, and recombination. 

The first studies including photoionization focused mainly on gas with a primordial composition. \citet{efstathiou_suppressing_1992} showed that an extragalactic UVB can strongly suppress the cooling of primordial gas at temperatures around $10^{4}$-$10^{5}$~K. Similar prescriptions were later incorporated into cosmological simulations, where optically thin primordial gas was assumed to be in ionization equilibrium with a uniform UVB \citep[e.g.][]{katz_cosmological_1996}. For chemically enriched gas, the problem becomes more complex because each element and ion responds differently to the gas density, temperature, and shape of the ionizing spectrum. A systematic implementation designed for cosmological simulations was presented by \citet{wiersma_effect_2009}, hereafter \wie. They used CLOUDY \citep[version 07.02;][]{Ferland98} to calculate cooling and heating rates for optically thin gas exposed to the CMB and the \citet[][hereafter HM01]{haardt_modelling_2001} UVB. Their tables depend on hydrogen number density, temperature, redshift, and chemical composition. The H and He contribution was tabulated for different helium abundances, while the contribution from heavy elements was calculated element by element. This allowed the total cooling rate to be reconstructed for non-solar abundance patterns using the contributions of individual elements.

\wie~ have shown that including photoionization can significantly reduce cooling by up to an order of magnitude, especially in diffuse gas with low density and in the temperature range of  $10^{4}\lesssim T\lesssim10^{6}$~K. The effect becomes weaker at higher densities, where collisional processes become more important, and at high temperatures, where the gas is already highly ionized. Their results showed that both heavy elements and photoionization are important for calculating the cooling of diffuse, chemically enriched gas.

The \wie~ cooling tables are still commonly used, while our description of the cosmic radiation field has evolved. Improved observational constraints on the cosmic star-formation history, galaxy UV emissivity, AGN populations, intergalactic absorption, and the timing of H~I and He~II reionization have led to updated UVB models, including those of \citet[][hereafter HM12]{haardt_radiative_2012}, \citet[][hereafter KS19]{khaire_new_2019}, and \citet[][hereafter FG20]{faucher-giguere_cosmic_2020}. Particularly at high redshift, where direct observational constraints are limited, the UVB, nevertheless, remains uncertain. These models differ in their assumptions about stellar spectra, escape fractions, AGN emissivities, absorber distributions, and reionization histories. As a result, they predict different intensities and spectral shapes of the ionizing background and therefore different ionization, cooling, and heating rates.

Several works have extended tabulated rates beyond optically thin gas exposed only to a uniform UVB. Within galaxies, gas is also photoionized by local sources such as young stars and AGN. Following \citet{Kannan2014}, \citet{Aura2019} included the local radiation field from stars and hot gas in the cooling and heating rates and found, in NIHAO zoom-in simulations, that it reduces stellar mass growth by $\sim20$ per cent and roughly halves the H\,{\sc i} mass by $z=0$, an effect stronger than that of changing the UVB model. Similarly, \citet{zhu2024} found in Auriga simulations of Milky Way-sized galaxies that radiation from young stars dominates over the UVB within the virial radius, suppressing star formation and the H\,{\sc i} content. \citet[][hereafter PS20]{ploeckingerSchaye2020} extended tabulated rates to the cold, dense interstellar medium by including an interstellar radiation field, self-shielding, cosmic rays, dust, and molecules, and \citet{Ploeckinger2025} coupled them to a non-equilibrium chemical network (\textsc{hybrid-chimes}). These developments improve the description of gas inside and around galaxies, but in the diffuse intergalactic gas the UVB remains the main ionizing source, and the cooling and heating rates depend directly on the adopted UVB model.

In this work, we provide an updated set of the element-by-element cooling and heating tables following the framework introduced by \wie. The new tables are computed in the same format as those of \wie, so that they can be directly used in codes that adopt the \wie~ tables, such as \textsc{OpenGadget3} \citep{OG3}, in which we release them (the complete table sets, in HDF5 format, are publicly available at \url{https://homepages.usm.uni-muenchen.de/OpenGadget/CoolingTables/}.). We calculate the new tables using the HM12, KS19, and FG20 UVB with the same numerical setup and abundance set, and we compare them with the calculations of \wie (HM01). The comparison with \wie~ quantifies the overall impact of updating the table calculations, including the radiation field, CLOUDY version 23.01, atomic data, and adopted abundance set. The comparison among HM12, KS19, and FG20 instead isolates the sensitivity to the adopted UVB within the new calculations.

\subsection{Comparison of the UVB models}

\begin{figure*}
    \centering
    \includegraphics[width=\textwidth]{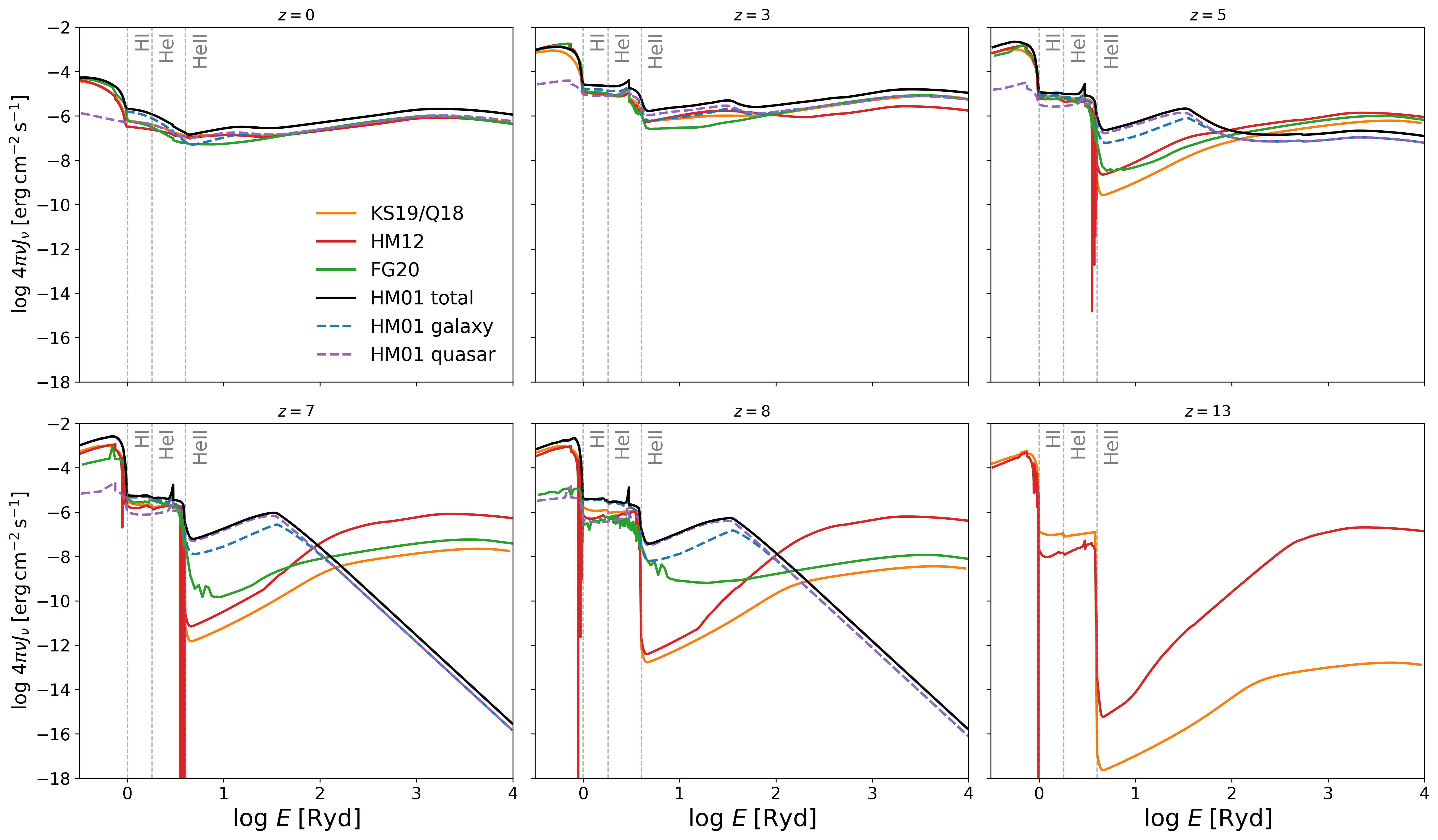}
    \caption{Comparison of the UVB spectra adopted in this work at $z=0$, $3$, $5$, $7$, $8$ and $13$.
    The quantity $4\pi J_\nu$ is shown as a function of photon energy in Rydbergs.
    The vertical dashed lines mark the ionization thresholds of H~I, He~I, and He~II at $1$, $1.81$, and $4$~Ryd, respectively.
    The input spectra for HM12 and KS19 were taken from the corresponding files distributed with CLOUDY 23.01.
    For comparison, we show the spectrum distributed with CLOUDY 07.02, the version used by W09, with the galaxy and quasar components shown separately, together with their sum. W09 refer to this background as HM01, although the corresponding file in that CLOUDY distribution is labelled HM05. The FG20 spectrum, which is not distributed with CLOUDY, was obtained from the publicly available UVB tables at \url{https://galaxies.northwestern.edu/uvb-fg20/}. The figure provides a direct view of the differences discussed below and shows how both the normalization and the hardness of the radiation field evolve with redshift. Only models available over the relevant redshift interval are displayed in each panel.}
    \label{fig:UVB_6panel_paper}
\end{figure*}

In this work, we construct net cooling tables using the HM12, FG20, and KS19 (fiducial Q18) UVB models. We also include HM01 in the following comparison because the cooling rates presented by  \wie, which were used as an external reference for our results, were calculated using this UVB model. The four models follow the same general approach, in which radiation produced by galaxies and AGN is propagated through the IGM. However, they differ in their assumptions about the source spectra, their redshift evolution, and the distribution of intervening absorbers. These differences affect both the intensity and spectral shape of the UVB and can therefore modify the ionization state, photoheating, and net cooling rates of the gas.
Each UVB model is described by its mean specific intensity $J_\nu$, i.e. the space- and angle-averaged monochromatic intensity of the background radiation, in units of erg~s$^{-1}$~cm$^{-2}$~Hz$^{-1}$~sr$^{-1}$ \citep{HM96, haardt_radiative_2012}. The quantity $4\pi J_\nu = c\,u_\nu$, where $u_\nu$ is the specific energy density of the radiation \citep{RL79}, directly sets the photoionization rate of each ion $i$,
\begin{equation}
\Gamma_i = \int_{\nu_i}^{\infty} \frac{4\pi J_\nu}{h\nu}\,\sigma_i(\nu)\,{\rm d}\nu ,
\end{equation}
where $h$ is the Planck constant, $\sigma_i(\nu)$ is the photoionization cross-section and $\nu_i$ the ionization threshold frequency, and, analogously, the corresponding photoheating rate \citep{haardt_radiative_2012}.
Figure~\ref{fig:UVB_6panel_paper} compares the spectral energy distributions of the four UVB models at selected redshifts. For HM01, the separate galaxy and quasar contributions are shown together with their sum.  

 The main differences among the UVB models considered in this work that are relevant to the present calculations are summarized in Table~\ref{tab:uvb_models}. A detailed description of the individual prescriptions is provided in Appendix~\ref{app:uvb_models}, where we also show their spectral shapes at different redshifts in Figure~\ref{fig:UVB_all_redshifts}.

\begin{table*}[p]
\centering
\caption{Summary of the main physical assumptions of the HM01, HM12, FG20, and KS19 UV/X-ray background models considered in this work. A detailed description of the individual prescriptions is provided in Appendix~\ref{app:uvb_models}.}
\label{tab:uvb_models}

\renewcommand{\arraystretch}{1.30}
\setlength{\tabcolsep}{5pt}

\begin{tabular*}{\textwidth}{
@{\extracolsep{\fill}}
p{0.145\textwidth}
p{0.195\textwidth}
p{0.195\textwidth}
p{0.195\textwidth}
p{0.195\textwidth}
@{}
}
\toprule
Component & \textcolor[HTML]{7F7F7F}{\textbf{HM01}} & \textcolor[HTML]{D62728}{\textbf{HM12}} & \textcolor[HTML]{2CA02C}{\textbf{FG20}} & \textcolor[HTML]{FF7F0E}{\textbf{KS19 (Q18)}} \\

\midrule

\textbf{Galaxy emissivity}
&
\citet{madau_high-redshift_1996} SFH; stellar emission normalized at 1500\,\AA; no independent SFRD reconstruction
&
SFRD(z) iteratively adjusted to reproduce dust-corrected FUV luminosity densities
&
Comoving emissivity measured at 1500~\AA;
\citealp{chiang_broadband_2019} at $z\leq2$;
UV LFs integrated to $M_{\rm UV}=-13$ at $z\geq2.1$
&
SFRD$(z)$ determined jointly with dust attenuation from multi-wavelength galaxy LFs
\\[2pt]

\midrule

\textbf{Stellar population}
&
YSFG template SED from Bruzual \& Charlot libraries;
assuming \citealp{salpeter_luminosity_1955} IMF;
Z=0.2 $Z_\odot$; 
and an age =0.5~Gyr
&
GALAXEV library \citep{bruzual_stellar_2003}; SSPs with $Z(z) = Z_\odot\,10^{-0.15z}$; \citet{salpeter_luminosity_1955} IMF, 0.1--100\,$M_\odot$
&
BPASS v2.2.1 including binaries;
$\alpha_1=1.30$ for $0.1$-$0.5\,M_\odot$,
$\alpha_2=2.45$ for $0.5$-$300\,M_\odot$;
continuous SF for 300~Myr;
$Z=0.1\,Z_\odot$
&
\textsc{Starburst99} \citealp{1999Leitherer};
\citealp{salpeter_luminosity_1955} IMF,
$0.1$-$100\,M_\odot$;
const $Z=0.008\simeq0.4\,Z_\odot$
\\[2pt]

\midrule

\textbf{Dust attenuation}
&
No explicit dust attenuation curve in the ionizing
galaxy emissivity
&
\citealp{calzetti_dust_2000} attenuation law;
$A_{\rm FUV}=1$~mag at $z\leq2$;
redshift-dependent $A_{\rm FUV}$ at higher $z$
&
\citealp{reddy_mosdef_2015}\ attenuation curve;
$E(B-V)=0.129$;
holes model
&
LMC2 extinction curve;
redshift-dependent $A_{\rm FUV}$
\\[2pt]

\midrule

\textbf{Ionizing escape fraction}
&
$f_{\rm esc}=0.1$
&
$f_{\rm esc}(z)=
1.8\times10^{-4}(1+z)^{3.4}$;
escaping stellar emission set to zero above 4~Ryd
&
$f_{\rm esc}(z)=
\min\!\left[
0.01\left(\frac{1+z}{4}\right)^{2.5},
0.2
\right]$;
no equivalent sharp cutoff above 4~Ryd
&
$f_{\rm esc}(z)=
\dfrac{10^{-5}}
{6.5\times10^{-5}+\exp(-2.4z)}$;
escaping stellar emission set to zero above 4~Ryd
\\[2pt]

\midrule

\textbf{AGN emissivity \& UV spectrum}
&
\citealp{boyle_2df_2000} optical QSO LF;
$L_\nu\propto\nu^{-1.8}$
below 1050~\AA
&
Fit to the 1\,Ryd emissivity from \citet{hopkins_observational_2007}; $L_\nu \propto \nu^{-0.44}$ ($\lambda > 1300$\,\AA), $\nu^{-1.57}$ ($\lambda < 1300$\,\AA) \citep{vandenberk2001, telfer2002}
&
\citealp{shen_bolometric_2020} bolometric LF;
Krawczyk et al.\ template;
$L_\nu\propto\nu^{-1.7}$
at 600-912~\AA
&
Observed type-1 QSO LF compilation;
free EUV spectral index;
fiducial $L_\nu\propto\nu^{-1.8}$
\\[2pt]

\midrule

\textbf{Obscured AGN \& X-ray spectrum}
&
Type-1 and synthetic type-2 AGN;
type-2 template from a separate Monte Carlo calculation
&
Hard X-ray LFs plus absorbing
H-column distribution;
obscured spectra reprocessed through absorption;
calibrated to the cosmic X-ray background
&
$f_{\rm obsc}=0.75$;
based on \citealp{sazonov_quasars_2004} obscured spectrum;
exponentially truncated X-ray power law
plus Compton reflection using PEXRAV
&
Type-1 and type-2 QSOs plus blazars;
modified \citealp{sazonov_quasars_2004} template;
spectrum extending from X-rays to TeV energies
\\[2pt]

\midrule

\textbf{IGM absorption}
&
single
power-law approximation to the column densities distribution
$f(N_{\rm HI})\propto N_{\rm HI}^{-1.5}$
with independently specified redshift evolution
&
Piecewise power law in $N_{\rm HI}$ and redshift;
five column-density regimes 
&
Modified \citealp{puchwein_consistent_2019} distribution;
redshift evolution changed to
$f(N_{\rm HI},z)\propto(1+z)$
at $z\leq1.56$
&
\citealp{inoue_updated_2014} analytical fitting function;
same prescription over the full redshift range
\\[2pt]

\midrule

\textbf{H\,I / He\,II reionization}
&
No specific calibrated H\,I or He\,II
reionization history
&
$z_{\rm rei,HI}=6.7$;
$z_{\rm rei,HeII}=2.8$
&
$z_{\rm rei,HI}=7.8$;
$z_{\rm rei,HeII}=3.5$
&
$z_{\rm rei,HI}=5.8$;
$z_{\rm rei,HeII}=2.8$
\\[2pt]

\midrule

\textbf{Electron-scattering optical depth}
&
--
&
$\tau_{\rm e}=0.084$
&
$\tau_{\rm e}=0.054$
&
$\tau_{\rm e}\simeq0.055$
\\

\bottomrule
\end{tabular*}

\end{table*}

%% METHOD

\section{Methodology}
\label{sec:method}
Our calculation follows the element-by-element method introduced by \wie. Radiative cooling and photoheating rates are computed with \textsc{CLOUDY 23.01} \citep{Cloudy23}. 
The section is organized as follows. Section~\ref{sec:cloudy} describes the physical setup of the CLOUDY calculations, the chemical elements included, and the set of CLOUDY runs and post-processing steps required to build the tables. Section~\ref{sec:radiative_cooling_and_heating} defines the normalized cooling, heating, and net cooling functions extracted from the CLOUDY output, which are the quantities stored in the tables, together with the cooling time. Section~\ref{sec:element_decomposition} explains how these functions are decomposed into the contribution of H and He and those of the individual heavy elements, and how the total rates can be reconstructed for an arbitrary abundance pattern. The grid in density-temperature, and the values of redshift on which all runs are performed are given in Section~\ref{sec:grid}. The runs for the H and He contribution at different helium mass fractions and the Solar reference composition used to normalize the metal contributions are described in Sections~\ref{sec:helium} and~\ref{sec:solar_abundances}, respectively. Finally, Section~\ref{sec:splitting} describes the separate storage of the cooling and heating rates and its possible use for an approximate treatment of shielding.

\subsection{CLOUDY calculations}
\label{sec:cloudy}

CLOUDY is a spectral synthesis code that, for a given chemical composition, density, temperature, and incident radiation field, determines the ionization state of the gas by solving the balance equations for all the ions included in its atomic database, and computes the resulting heating and cooling rates \citep{Cloudy23}.

In our calculations, the gas is in ionization equilibrium at a fixed temperature (using \texttt{constant temperature} command). The incident radiation field consists of the CMB at the corresponding redshift and one of the HM12, FG20, or KS19 UVB models. Each model consists of a single, geometrically thin zone (\texttt{set dr 0} and \texttt{stop zone 1}), so that the incident radiation is not attenuated and the gas is effectively optically thin. We therefore neglect local self-shielding. Induced processes are disabled (via \texttt{no induced processes} command) and, consistently with \wie, dust processes are excluded.

We include the 11 elements considered by \wie, namely H, He, C, N, O, Ne, Mg, Si, S, Ca, and Fe, and extend the chemical network by adding Na, Al, Ar, and Ni. The complete set of tables therefore contains 15 chemical elements, whose adopted Solar reference abundances are described in Section~\ref{sec:solar_abundances}. The contribution of each element to heating and cooling is stored separately, so that users can select the elements available in their simulation or chemical model (Section~\ref{sec:element_decomposition})

For each UVB model and redshift, the tables require CLOUDY runs for 26 chemical compositions, whose role is detailed in the following sections. These are: the full CLOUDY Solar mixture and 13 compositions in which one heavy element at a time is removed, used to derive the contributions of the individual elements (Section~\ref{sec:element_decomposition}); a reduced Solar composition limited to the 15 tracked elements, which provides the Solar rates, electron abundance, and mean particle mass stored in the tables; and 11 metal-free compositions consisting exclusively of H and He with different helium mass fractions (Section~\ref{sec:helium}). 

The rates for each composition are computed on the density-temperature grid described in Section~\ref{sec:grid}, generated with the CLOUDY \texttt{grid} command, which enables running the individual models in parallel. This corresponds to $81\times352\times26 = 741\,312$ CLOUDY models per redshift, and to $\approx 1.33\times10^{8}$ models for the full set of tables. From each model, we extract the total heating and cooling rates (\texttt{save cooling}) and, for the Solar and metal-free (H and He) runs, the quantities needed to derive the electron abundance, the mean particle mass, and the hydrogen fractions (\texttt{save overview}). These outputs are then post-processed to compute the quantities defined in Sections~\ref{sec:radiative_cooling_and_heating} and~\ref{sec:element_decomposition}, which are written as the HDF5 tables described in Section~\ref{sec:tables}.

%Following Ildar's suggestion:
\subsection{Radiative cooling and heating rates}
\label{sec:radiative_cooling_and_heating}

CLOUDY computes the total volumetric heating and cooling rates, which we denote by $\mathcal{H}$ and $\mathcal{C}$, respectively. 
Both quantities are expressed in $\mathrm{erg\,cm^{-3}\,s^{-1}}$ and are extracted from the \texttt{Htot} and \texttt{Ctot} columns of the CLOUDY \texttt{save cooling} outputs.

During post-processing, we divide both rates by the square of the hydrogen number density, $n_{\rm H}$, and define:
\begin{equation}
\Lambda_{\rm cool} = \frac{\mathcal{C}}{n_{\rm H}^{2}},
\qquad
\Gamma_{\rm heat} = \frac{\mathcal{H}}{n_{\rm H}^{2}}.
\label{eq:normalized_cooling_heating}
\end{equation}
Both functions have units of $\mathrm{erg\,cm^{3}\,s^{-1}}$. The normalized net cooling function is
\begin{equation}
\Lambda_{\rm net}
= \Lambda_{\rm cool} - \Gamma_{\rm heat}
= \frac{\mathcal{C}-\mathcal{H}}{n_{\rm H}^{2}}.
\label{eq:net_cooling_definition}
\end{equation}

With this sign convention, $\Lambda_{\rm net}>0$ corresponds to net cooling, $\Lambda_{\rm net}<0$ to net heating, and $\Lambda_{\rm net}=0$ to thermal balance. 

For an ideal monatomic gas, the internal energy density (per unit volume) is

\begin{equation}
    u_{\rm th}(n_{\rm H},T,z)
    =
    \frac{3}{2}\,
    n_{\rm tot}(n_{\rm H},T,z)\,
    k_{\rm B}T,
    \label{eq:uth}
\end{equation}
where $n_{tot}$ is the total number density of particles and $k_{\rm B}$ is the Boltzmann constant. We define the cooling time as the ratio of the thermal energy density to the volumetric energy loss rate:

\begin{equation}
t_{\rm cool} =
\frac{u_{\rm th}}{n_{\rm H}^{2}\Lambda_{\rm net}},
\qquad \Lambda_{\rm net}>0.
\label{eq:cooling_time_general}
\end{equation}

The cooling time is therefore defined only for gas undergoing net radiative cooling. The calculation of $t_{\rm cool}$ using the quantities provided in our cooling tables is described in Sect.~\ref{sec:results_cooling_times}.

\subsection{Element-by-element decomposition}
\label{sec:element_decomposition}

 The datasets stored in the tables are meant to be used to reconstruct the normalized net cooling rate as the sum of the H and He contribution and the contributions of individual heavy elements. For an arbitrary abundance pattern, it can be approximated as
\begin{equation}
    \Lambda_{\rm net}
    =
    \Lambda_{\rm net,HHe}
    +
    \sum_{i>{\rm He}}
    \Lambda_{{\rm net},i,\odot}
    \left[
        \frac{(n_i/n_{\rm H})}
             {(n_i/n_{\rm H})_\odot}
    \right]
    \left[
        \frac{n_e/n_{\rm H}}
             {(n_e/n_{\rm H})_\odot}
    \right],
    \label{eq:cooling_sum}
\end{equation}
where $\Lambda_{\rm net,HHe}$ is the normalized net cooling rate from H and He, $\Lambda_{{\rm net},i,\odot}$ is the normalized net cooling rate associated with element $i$ for the adopted Solar reference composition, $n_i/n_{\rm H}$ is the number abundance of element $i$ relative to hydrogen, and $n_e/n_{\rm H}$ is the electron abundance, with $n_e$ the free-electron number density. The electron abundance $n_e/n_{\rm H}$ is taken from the H and He tables at the helium mass fraction of the gas, while $(n_e/n_{\rm H})_\odot$ is taken from the Solar tables. If the He to H abundance ratio is fixed, the electron abundance correction factor is equal to unity with an accuracy of a few percent.

For each heavy element, the Solar-abundance contribution is obtained from two almost identical CLOUDY calculations: the first one includes all the elements of the full CLOUDY Solar mixture, whereas in the second one, element $i$ is removed (Section~\ref{sec:cloudy}). The difference between the two normalized net cooling rates defines the contribution of that element:
\begin{equation}
    \Lambda_{{\rm net},i,\odot}
    =
    \Lambda_{{\rm net},\odot}
    -
    \Lambda_{{\rm net},\odot,\,i=0}.
    \label{eq:element_difference}
\end{equation}

This approximation assumes that heavy elements do not provide a significant fraction of free electrons and that their cooling contributions scale linearly with their abundances. The correction involving $n_e/n_{\rm H}$ accounts separately for variations in the helium abundance, which can modify the electron density and consequently the efficiency of collisional cooling (\wie).

\subsection{Density, temperature, and redshift}
\label{sec:grid}

The rates are tabulated as functions of  $n_{\rm H}$ and  $T$, for fixed values of redshift $z$. We adopt the same density and temperature resolution as \wie:
\begin{align}
    -8 &\leq
    \log_{10}\left(\frac{n_{\rm H}}{{\rm cm}^{-3}}\right)
    \leq 0,
    &
    \Delta\log_{10} n_{\rm H} &= 0.1,
    \label{eq:density_grid}\\
    2 &\leq
    \log_{10}\left(\frac{T}{\rm K}\right)
    \leq 8.982,
    &
    \Delta\log_{10} T &= 0.0199.
    \label{eq:temperature_grid}
\end{align}
The quantities $\Delta\log_{10}n_{\rm H}$ and $\Delta\log_{10}T$ denote the logarithmic spacing in the $\mathrm{n_H}$ and temperature grids, respectively. The grid contains 81 gas density values and 352 temperature values. The gas density bins span from $10^{-8}\,\mathrm{cm}^{-3}$ to $1\,\mathrm{cm}^{-3}$, while the temperature bins span from $10^{2}\,\mathrm{K}$ to $9.6\times10^{8}\,\mathrm{K}$, consistently with the temperature sampling of the \wie~ tables.

Up to $z = 8.989$, we adopt the same redshift values as \wie, which are approximately uniformly spaced in $\log_{10}(1+z)$, with $\Delta\log_{10}(1+z)\simeq0.02$. At higher redshifts, we extend each table set up to the maximum redshift available for the corresponding UVB model, with a spacing of $\Delta z \simeq 0.2$--$0.7$. This results in 61 redshift values over $0\leq z\leq15.1$ for HM12, 67 values over $0\leq z\leq14.898$ for KS19, and 51 values over $0\leq z\leq9.7$ for FG20. 

\subsection{Treatment of hydrogen and helium}
\label{sec:helium}

The combined H and He contribution is calculated from CLOUDY runs in which the gas is composed exclusively of H and He, with a specific He mass fraction adopted for each run. These calculations are repeated for several He mass fractions so that the rates can be applied to gas with a non-primordial He fraction.

We adopt 11 values for the He mass fraction in the range $0.238\leq Y\leq0.338$, with a linear spacing of $\Delta Y = 0.01$, where
\begin{equation}
    Y = \frac{M_{\rm He}}{M_{\rm H}+M_{\rm He}}.
\end{equation}
Here, $M_{\rm H}$ and $M_{\rm He}$ are the H and He masses, respectively. The corresponding He number ratio is
\begin{equation}
    \frac{n_{\rm He}}{n_{\rm H}}
    =
    \frac{m_{\rm H}}{m_{\rm He}}\,\frac{Y}{1-Y},
    \label{eq:helium_number_ratio}
\end{equation}

Here, $n_{\rm He}$ is the helium number density. The adopted values are listed in Table~\ref{tab:helium}. The first seven values reproduce the range considered by \wie, while the final four extend the tables to more He-rich gas.

Each value of $Y$ is implemented in a separate CLOUDY run through the \texttt{element helium abundance} command. For each run, we tabulate the cooling and heating rates from H and He, together with the electron abundance $n_e/n_{\rm H}$, the mean particle mass, and the hydrogen fractions (Section~\ref{sec:tables}).

\begin{table}[t]
\centering
\caption{Helium mass fractions and corresponding He-to-hydrogen number ratios adopted for the H and He tables. The values below the horizontal line extend the range considered by \wie.}
\label{tab:helium}
\begin{tabular}{cc}
\toprule
$Y$ & $n_{\rm He}/n_{\rm H}$ \\
\midrule
0.238 & 0.078653 \\
0.248 & 0.083047 \\
0.258 & 0.087560 \\
0.268 & 0.092197 \\
0.278 & 0.096962 \\
0.288 & 0.101860 \\
0.298 & 0.106898 \\
\midrule
0.308 & 0.112070 \\
0.318 & 0.117400 \\
0.328 & 0.122900 \\
0.338 & 0.128560 \\
\bottomrule
\end{tabular}
\end{table}

\subsection{Solar reference abundances}
\label{sec:solar_abundances}

The metal contributions are normalized to the CLOUDY GASS10 Solar composition \citep{Grevesse2010}, implemented through \texttt{solar\_GASS10.abn}. For each element $i$, Table~\ref{tab:solar_abundances} lists the Solar number abundance relative to hydrogen, $(n_i/n_{\rm H})_\odot$, and the corresponding Solar mass fraction, $X_{i,\odot}$. Both quantities are stored in the header of each table file. The mass fractions are computed over the 15 elements tracked in the tables, excluding the trace elements present in the full CLOUDY Solar mixture.

\begin{table}[t]
\centering
\caption{Solar reference abundances stored in the HDF5 files. The second column gives the number abundance relative to hydrogen,
$(n_i/n_{\rm H})_\odot$, and the third gives the corresponding mass fraction stored as \texttt{Solar\_mass\_fractions}. The adopted number abundances follow the CLOUDY GASS10 solar composition.}
\label{tab:solar_abundances}
\begin{tabular}{lcc}
\toprule
Element & $(n_i/n_{\rm H})_\odot$ & $X_{i,\odot}$ \\
\midrule
H  & $1.00$              & $0.737$ \\
He & $8.51\times10^{-2}$ & $0.249$ \\
C  & $2.69\times10^{-4}$ & $2.36\times10^{-3}$ \\
N  & $6.76\times10^{-5}$ & $6.93\times10^{-4}$ \\
O  & $4.90\times10^{-4}$ & $5.74\times10^{-3}$ \\
Ne & $8.51\times10^{-5}$ & $1.26\times10^{-3}$ \\
Na & $1.74\times10^{-6}$ & $2.93\times10^{-5}$ \\
Mg & $3.98\times10^{-5}$ & $7.08\times10^{-4}$ \\
Al & $2.82\times10^{-6}$ & $5.57\times10^{-5}$ \\
Si & $3.24\times10^{-5}$ & $6.66\times10^{-4}$ \\
S  & $1.32\times10^{-5}$ & $3.10\times10^{-4}$ \\
Ar & $2.51\times10^{-6}$ & $7.34\times10^{-5}$ \\
Ca & $2.19\times10^{-6}$ & $6.42\times10^{-5}$ \\
Fe & $3.16\times10^{-5}$ & $1.29\times10^{-3}$ \\
Ni & $1.66\times10^{-6}$ & $7.13\times10^{-5}$ \\
\bottomrule
\end{tabular}
\end{table}

\subsection{Separate storage of cooling and heating rates}
\label{sec:splitting}

In addition to the net cooling function, we store the normalized cooling and heating functions separately at each grid point (Eqs.~\ref{eq:normalized_cooling_heating} and~\ref{eq:net_cooling_definition}), whereas the tables of \wie~ primarily provide the net rate. This separation also makes it possible in principle to attenuate the photoheating term independently, providing an approximate treatment of radiative shielding. Rather than treating the attenuation of individual ionizing species separately, a single effective shielding factor can be applied to the combined photoheating contribution from H~I, He~I, and He~II. This should be regarded as a mean shielding prescription rather than a self-consistent treatment.

The element-by-element decomposition of Section~\ref{sec:element_decomposition} applies consistently to the cooling, heating, and net rates.

% TABLE DESCRIPTION

\section{Table Description}
\label{sec:tables}

In this section, we describe the structure and content of the released cooling tables. We first present the file format, naming convention, and size of the table sets (Section~\ref{sec:format}), and then list the datasets stored in each file (Section~\ref{sec:content}).

\subsection{File format}
\label{sec:format}

All tables are stored in HDF5 format. Each HDF5 file corresponds to a single redshift slice and contains all the quantities listed in Section~\ref{sec:content}. The file naming convention is \texttt{<UVB>/z\_<redshift>.hdf5}, where \texttt{<UVB>} is one of \texttt{HM12}, \texttt{KS19}, or \texttt{FG20}, and \texttt{<redshift>} is the redshift value with three decimal places (e.g., \texttt{KS19/z\_0.000.hdf5}). Each file follows and extends the hierarchical structure of the \wie~ tables, to ensure compatibility with existing simulation codes.

The three table sets are released as separate archives, each containing one file for each of the redshift values listed in Section~\ref{sec:grid}. The complete release occupies approximately 3.2 GB of storage. The individual FG20, HM12, and KS19 table sets require approximately 0.92, 1.10, and 1.21 GB, respectively, reflecting primarily the different numbers of redshift slices.

\subsection{Table contents}
\label{sec:content}

Table~\ref{tab:contents} summarizes the datasets stored in each HDF5 file. For each dataset, the table gives its full path within the file, its dimensions, and a short description of its content, including the physical units where applicable. The axis arrays ($n_{\rm H}$, $T$, $Y$) are stored once per file and shared across all datasets.

\begin{table*}[h!]
\centering
\caption{Datasets stored in each HDF5 redshift file. The grid dimensions are $N_{n_{\rm H}} = 81$ ($n_{\rm H}$ bins), $N_T = 352$ ($T$ bins), and $N_Y = 11$ helium mass fraction values. The header lists the $N_{\rm el} = 13$ tracked elements and the $N_s = 13$ individually tabulated heavy elements. Cooling and heating rates are normalized by $n_{\rm H}^2$ (Section~\ref{sec:radiative_cooling_and_heating}), in units of erg\,cm$^3$\,s$^{-1}$.}
\label{tab:contents}
\setlength{\tabcolsep}{24pt}
\begin{tabular*}{\tblwidth}{@{}L@{\hspace{20pt}}L@{\hspace{26pt}}L@{}}
\toprule
Dataset / Group & Dimensions & Description \\
\midrule
\multicolumn{3}{l}{\textit{Header group}} \\
\texttt{Header/Redshift}                      & (1)     & Redshift of this file \\
\texttt{Header/Number\_of\_density\_bins}     & scalar  & $N_{n_{\rm H}}$ \\
\texttt{Header/Number\_of\_temperature\_bins} & scalar  & $N_T$ \\
\texttt{Header/Number\_of\_helium\_fractions} & scalar  & $N_Y$ \\
\texttt{Header/Number\_of\_species}           & scalar  & $N_s$ \\
\texttt{Header/Species\_names}                & ($N_s$) & Names of the individually tabulated heavy elements \\
\textit{Header/Abundances} \\
\ \ \ \texttt{Header/Abundances/Abund\_names}           & ($N_{\rm el}$) & Names of the tracked elements \\
\ \ \ \texttt{Header/Abundances/Number\_of\_abundances} & scalar         & $N_{\rm el}$ \\
\ \ \ \texttt{Header/Abundances/Solar\_mass\_fractions} & ($N_{\rm el}$) & $X_{i,\odot}$ \\
\ \ \ \texttt{Header/Abundances/Solar\_number\_ratios}  & ($N_{\rm el}$) & $(n_i/n_{\rm H})_\odot$ \\
\midrule
\multicolumn{3}{l}{\textit{Metal\_free group}} \\
\texttt{Metal\_free/Hydrogen\_density\_bins}      & ($N_{n_{\rm H}}$) & $n_{\rm H}\ [{\rm cm}^{-3}]$ \\
\texttt{Metal\_free/Temperature\_bins}            & ($N_T$)           & $T\ [{\rm K}]$ \\
\texttt{Metal\_free/Helium\_mass\_fraction\_bins} & ($N_Y$)           & $Y$ \\
\texttt{Metal\_free/Helium\_number\_ratio\_bins}  & ($N_Y$)           & $n_{\rm He}/n_{\rm H}$ \\
\texttt{Metal\_free/Net\_Cooling}  & ($N_Y$, $N_T$, $N_{n_{\rm H}}$) & $\Lambda_{\rm net,HHe}$ [erg\,cm$^3$\,s$^{-1}$] \\
\texttt{Metal\_free/Cooling\_only} & ($N_Y$, $N_T$, $N_{n_{\rm H}}$) & $\Lambda_{\rm cool,HHe}$ [erg\,cm$^3$\,s$^{-1}$] \\
\texttt{Metal\_free/Heating\_only} & ($N_Y$, $N_T$, $N_{n_{\rm H}}$) & $\Gamma_{\rm heat,HHe}$ [erg\,cm$^3$\,s$^{-1}$] \\
\texttt{Metal\_free/Electron\_density\_over\_n\_h}& ($N_Y$, $N_T$, $N_{n_{\rm H}}$) & $n_e/n_{\rm H}$ \\
\texttt{Metal\_free/Mean\_particle\_mass}         & ($N_Y$, $N_T$, $N_{n_{\rm H}}$) & Mean particle mass $\mu$ [$m_{\rm H}$] \\
\textit{Metal\_free/Hydrogen\_fractions} \\
\ \ \ \texttt{Metal\_free/Hydrogen\_fractions/H2}  & ($N_Y$, $N_T$, $N_{n_{\rm H}}$) & $x_{\rm H_2}=2n_{\rm H_2}/n_{\rm H}$ \\
\ \ \ \texttt{Metal\_free/Hydrogen\_fractions/HI}  & ($N_Y$, $N_T$, $N_{n_{\rm H}}$) & $x_{\rm HI}=n_{\rm HI}/n_{\rm H}$ \\
\ \ \ \texttt{Metal\_free/Hydrogen\_fractions/HII} & ($N_Y$, $N_T$, $N_{n_{\rm H}}$) & $x_{\rm HII}=n_{\rm HII}/n_{\rm H}$ \\
\textit{Metal\_free/Temperature} \\
\ \ \ \texttt{Metal\_free/Temperature/Hydrogen\_density\_bins}      & ($N_{n_{\rm H}}$) & $n_{\rm H}\ [{\rm cm}^{-3}]$ \\
\ \ \ \texttt{Metal\_free/Temperature/Helium\_mass\_fraction\_bins} & ($N_Y$)           & $Y$ \\
\ \ \ \texttt{Metal\_free/Temperature/Helium\_number\_ratio\_bins}  & ($N_Y$)           & $n_{\rm He}/n_{\rm H}$ \\
\ \ \ \texttt{Metal\_free/Temperature/Energy\_density\_bins}        & ($N_T$)           & Specific internal energy $u\ [{\rm erg\,g}^{-1}]$ \\
\ \ \ \texttt{Metal\_free/Temperature/Temperature}                  & ($N_Y$, $N_T$, $N_{n_{\rm H}}$) & $T(u,\,n_{\rm H},\,Y)\ [{\rm K}]$ \\
\midrule
\multicolumn{3}{l}{\textit{Solar group (Solar composition restricted to the 15 tracked elements)}} \\
\texttt{Solar/Hydrogen\_density\_bins}       & ($N_{n_{\rm H}}$)        & $n_{\rm H}\ [{\rm cm}^{-3}]$ \\
\texttt{Solar/Temperature\_bins}             & ($N_T$)                  & $T\ [{\rm K}]$ \\
\texttt{Solar/Net\_Cooling}                  & ($N_T$, $N_{n_{\rm H}}$) & Net cooling [erg\,cm$^3$\,s$^{-1}$] \\
\texttt{Solar/Cooling\_only}                 & ($N_T$, $N_{n_{\rm H}}$) & Cooling [erg\,cm$^3$\,s$^{-1}$] \\
\texttt{Solar/Heating\_only}                 & ($N_T$, $N_{n_{\rm H}}$) & Heating [erg\,cm$^3$\,s$^{-1}$] \\
\texttt{Solar/Electron\_density\_over\_n\_h} & ($N_T$, $N_{n_{\rm H}}$) & $(n_e/n_{\rm H})_\odot$ \\
\texttt{Solar/Mean\_particle\_mass}          & ($N_T$, $N_{n_{\rm H}}$) & Mean particle mass $\mu$ [$m_{\rm H}$] \\
\midrule
\multicolumn{3}{l}{\textit{Individual element groups (\texttt{<El>} $\in$ \{Aluminium, Argon, Calcium, Carbon, Iron, Magnesium, Neon, Nickel, Nitrogen, Oxygen,}} \\
\multicolumn{3}{l}{\textit{Silicon, Sodium, Sulphur\})}} \\
\texttt{<El>/Hydrogen\_density\_bins} & ($N_{n_{\rm H}}$)        & $n_{\rm H}\ [{\rm cm}^{-3}]$ \\
\texttt{<El>/Temperature\_bins}       & ($N_T$)                  & $T\ [{\rm K}]$ \\
\texttt{<El>/Net\_Cooling}            & ($N_T$, $N_{n_{\rm H}}$) & $\Lambda_{{\rm net},i,\odot}$ [erg\,cm$^3$\,s$^{-1}$] \\
\texttt{<El>/Cooling\_only}           & ($N_T$, $N_{n_{\rm H}}$) & $\Lambda_{{\rm cool},i,\odot}$ [erg\,cm$^3$\,s$^{-1}$] \\
\texttt{<El>/Heating\_only}           & ($N_T$, $N_{n_{\rm H}}$) & $\Gamma_{{\rm heat},i,\odot}$ [erg\,cm$^3$\,s$^{-1}$] \\
\midrule
\multicolumn{3}{l}{\textit{Total\_Metals group (sum of the 13 element contributions)}} \\
\texttt{Total\_Metals/Hydrogen\_density\_bins} & ($N_{n_{\rm H}}$)        & $n_{\rm H}\ [{\rm cm}^{-3}]$ \\
\texttt{Total\_Metals/Temperature\_bins}       & ($N_T$)                  & $T\ [{\rm K}]$ \\
\texttt{Total\_Metals/Net\_Cooling}            & ($N_T$, $N_{n_{\rm H}}$) & $\sum_i \Lambda_{{\rm net},i,\odot}$ [erg\,cm$^3$\,s$^{-1}$] \\
\texttt{Total\_Metals/Cooling\_only}           & ($N_T$, $N_{n_{\rm H}}$) & $\sum_i \Lambda_{{\rm cool},i,\odot}$ [erg\,cm$^3$\,s$^{-1}$] \\
\texttt{Total\_Metals/Heating\_only}           & ($N_T$, $N_{n_{\rm H}}$) & $\sum_i \Gamma_{{\rm heat},i,\odot}$ [erg\,cm$^3$\,s$^{-1}$] \\
\bottomrule
\end{tabular*}
\end{table*}

All quantities, including the density, temperature, and specific-internal-energy axis arrays, are stored as linear values. Following the CLOUDY convention, the molecular fraction $x_{\rm H_2}$ counts the hydrogen atoms bound in molecules, so that $x_{\rm HI}+x_{\rm HII}+x_{\rm H_2}=1$.

% RESULTS

\section{Results}
\label{sec:results}

In this section, we examine how the choice of UV/X-ray background affects the thermal rates and the ionization state predicted by the CLOUDY calculations. We first compare the net cooling rates and the resulting thermal-equilibrium temperatures. We then use the tabulated hydrogen fractions and the element-by-element cooling contributions to identify the physical origin of the differences. Finally, we discuss how these changes propagate into the net cooling time and into the regions where radiative cooling is effective on cosmological or gravitational timescales.

Throughout the main analysis, we adopt the Solar composition at all redshifts. This choice should not be interpreted as an assumption that the gas is expected to have a Solar composition, particularly at high redshift. Rather, Solar composition is used as a fixed reference composition, held constant across the full redshift range in order to provide a homogeneous basis for comparing the different UVB prescriptions and their redshift evolution. The corresponding calculations that include only H and He are collected in the Appendix~\ref{app:hhe_results}, where we show that the main qualitative UVB-dependent trends persist when metal-line cooling is removed.

As discussed in the Introduction, the comparison with \wie~ represents the combined effect of updating the table calculations, whereas comparisons between KS19, HM12, and FG20 isolate the dependence on the adopted UVB within the new calculations.

\subsection{Evolution of the net cooling rates}
\label{sec:results_net_rates}

\begin{figure*}
    \centering
    \includegraphics[width=\textwidth]
    {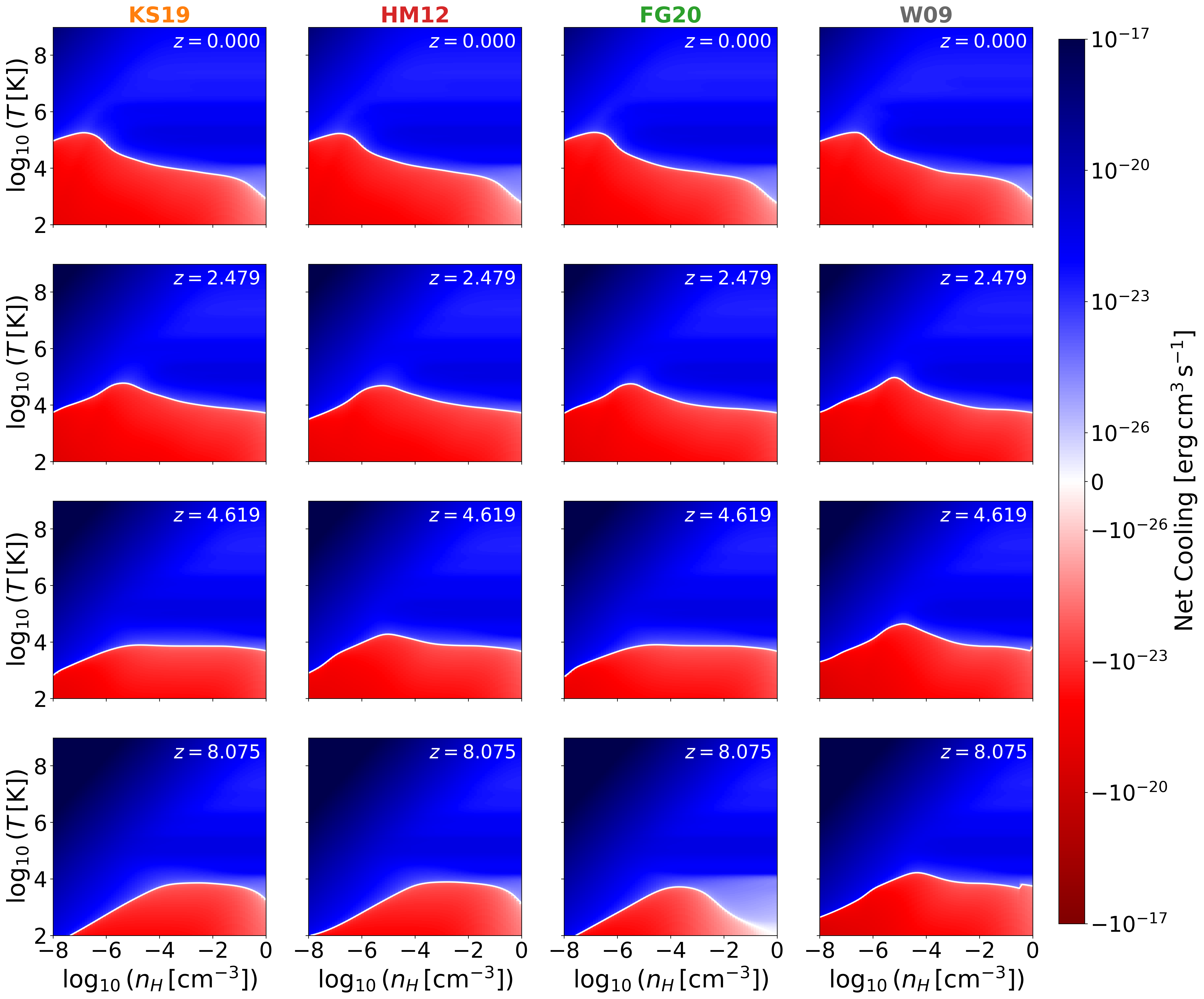}
    \caption{
    Net cooling efficiency in the density-temperature plane for a fixed
    Solar reference composition at $z=0.000$, $2.479$, $4.619$, and
    $8.075$. Columns show KS19, HM12, FG20, and \wie~ (HM01). Positive
    values correspond to net cooling and negative values to net heating.
    The white curve marks thermal equilibrium,
    $\Lambda_{\rm net}=0$.
    }
    \label{fig:net_cooling_maps}
\end{figure*}

Figure~\ref{fig:net_cooling_maps} shows the net cooling function $\Lambda_{\rm net}$ (Eq.~\ref{eq:net_cooling_definition}) obtained with the different UVB prescriptions at $z=0.000$, $2.479$, $4.619$, and $8.075$, using the Solar reference calculations adopted throughout the main analysis. Besides $z=0$, among all the computed redshifts, these values were selected because they illustrate the regimes in which the differences between the UVB models are most evident. The comparison at all redshifts shows the same qualitative trends. In each panel, the white curve marks $\Lambda_{\rm net}=0$, i.e. thermal equilibrium, which separates the net-heating and net-cooling regimes. 

As shown in Figure~\ref{fig:net_cooling_maps}, at $z=0$, all four prescriptions produce the same characteristics. Gas below the equilibrium temperature is dominated by net heating, whereas gas above it undergoes net cooling. The models converge increasingly towards high temperatures and high densities, where collisional processes and free-free emission become progressively more important relative to photoionization and photoheating. The largest residual differences are therefore found in diffuse gas and close to the transition between the net-heating and net-cooling regimes.

At $z=2.479$, differences between the models become more apparent, particularly around the low-density equilibrium boundary. At these densities, the thermal state of the gas is sensitive to the intensity and spectral shape of the ionizing background. Differences among the UVB prescriptions consequently affect both the photoheating rate and the ionization balance of the gas, which in turn modifies the radiative cooling efficiency.

The differences become more pronounced at $z=4.619$. \wie~ shows a distinct transition between heating and cooling compared with the more recent calculations. Among the three new table sets, HM12 also departs from KS19 and FG20 over part of the diffuse regime, while KS19 and FG20 remain comparatively similar over much of the density-temperature plane.

The strongest model dependence among the considered redshifts is found at $z=8.075$. In particular, FG20 develops a very different low-temperature structure over a limited density interval. At these redshifts, the UVB prescriptions differ substantially because of their different assumptions regarding the evolution of the ionizing background and reionization. These differences are reflected most clearly in cold, diffuse, photoionized gas, where the thermal balance is particularly sensitive to the radiation field. By contrast, the models progressively converge towards the hot, collisionally ionized regime, where the detailed properties of the UVB have a smaller influence on the net cooling rate.

\subsection{Thermal-equilibrium temperatures}
\label{sec:results_equilibrium}

For a given $n_{\rm H}$, redshift and chemical composition, the thermal-equilibrium temperature, $T_{\rm eq}$, is defined by the
condition

\begin{equation}
    \Lambda_{\rm net}(n_{\rm H},T_{\rm eq},z)=0,
    \label{eq:thermal_equilibrium}
\end{equation}

where $T_{\rm eq}$ is the equilibrium temperature at which radiative cooling and heating exactly balance. 

\begin{figure*}
    \centering
    \includegraphics[width=\textwidth]
    {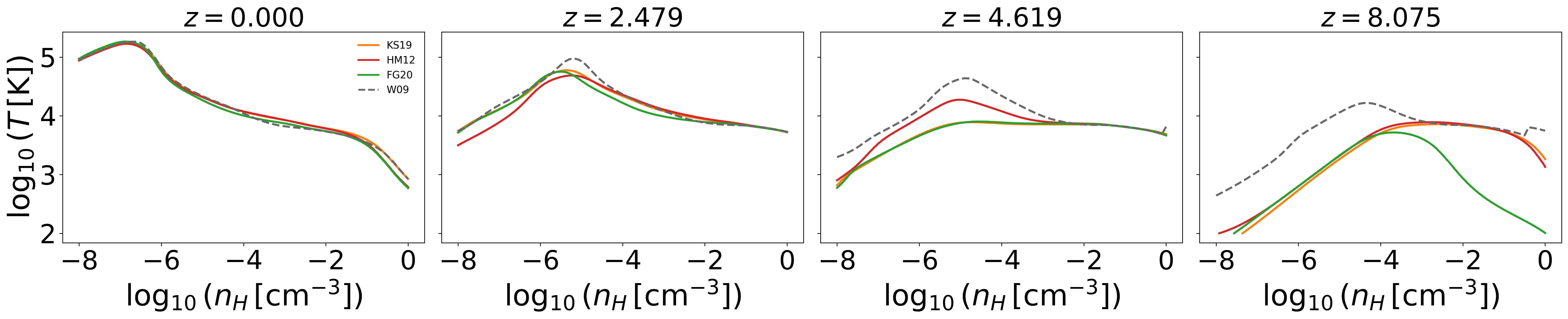}
    \caption{
    Thermal-equilibrium points as a function of hydrogen number density for
    KS19, HM12, FG20, and \wie~ (HM01), at $z=0.000$, $2.479$, $4.619$,
    and $8.075$. The curves correspond to the $\Lambda_{\rm net}=0$
    contours shown in Figure~\ref{fig:net_cooling_maps}, here overlaid to
    facilitate a direct comparison among the models. KS19, HM12, and FG20
    use the same adopted Solar reference composition, whereas \wie~ retains
    the solar abundance set of the original calculations.}
    \label{fig:thermal_equilibrium}
\end{figure*}

Figure~\ref{fig:thermal_equilibrium} provides a direct comparison of the thermal-equilibrium points identified in the net cooling maps shown in Section~\ref{sec:results_net_rates}. Specifically, the $\Lambda_{\rm net}=0$ contours shown in Figure~\ref{fig:net_cooling_maps} are extracted and overlaid as a function of$n_{\rm H}$. This representation makes differences among the models, as well as their evolution with redshift, easier to identify than in the density-temperature maps.

At $z=0$, the equilibrium curves are very close over most of the density range. Quantitatively, over the diffuse-gas interval $-8 \leq \log_{10}(n_{\rm H}/{\rm cm}^{-3}) \leq -3$, the median difference in $\log_{10}T_{\rm eq}$ between \wie~ and each of the three new calculations is smaller than $0.05$ dex. The different calculations therefore predict very similar equilibrium temperatures at the present epoch.

At $z=2.479$, the models begin to separate more clearly. Over the same diffuse-density interval, the median offset of \wie~ relative to the three new calculations ranges from approximately $0.04$ to $0.17$ dex, while local differences reach approximately $0.3$ dex. The separation is not uniform with density, but is most apparent over the density ranges where the equilibrium boundary is particularly sensitive to the adopted radiation field.

The differences become substantially larger at $z=4.619$. \wie~ lies above the three new calculations throughout the comparable diffuse-density interval. Its median equilibrium-temperature excess ranges from approximately $0.24$ to $0.42$ dex, corresponding to factors of approximately $1.7$-$2.6$ in $T_{\rm eq}$. The difference is strongly density-dependent and locally reaches approximately $0.75$ dex, or a factor of $\sim5.6$ in equilibrium temperature. Among the three new table sets, HM12 also shows a higher-temperature structure over part of the diffuse regime relative to KS19 and FG20.

The strongest discrepancy occurs at $z=8.075$. \wie~ predicts higher equilibrium temperatures than all three new calculations throughout the comparable diffuse-density interval. The median offset is $0.72$-$0.82$ dex, corresponding to equilibrium temperatures higher by factors of approximately $5$-$7$. The largest difference approaches $0.9$ dex, or nearly a factor of $8$ in temperature. 

\subsection{Hydrogen fractions}
\label{sec:results_hydrogen}

The tabulated hydrogen fractions $x_{\rm HI}$, $x_{\rm HII}$, and $x_{\rm H_2}$ (Section~\ref{sec:content}) provide a direct link between the incident UVB and the thermal differences described above. They correspond to the hydrogen fractions returned by CLOUDY for each combination of helium abundance, density, and temperature in the H and He calculations. Hydrogen fractions are not available in the \wie~ tables considered here and are therefore not included in this comparison.

The tabulated fractions should be interpreted within the physical assumptions of the CLOUDY calculations. In particular, while $x_{\rm HI}$ and $x_{\rm HII}$ directly trace the ionization state of the optically thin gas considered here, the interpretation of $x_{\rm H_2}$ requires additional caution. The calculations are dust-free and optically thin and do not include attenuation of the incident UVB by local self-shielding. They also exclude local interstellar radiation fields, cosmic rays, and non-equilibrium chemistry. Since these processes can affect the formation and destruction of molecular hydrogen, particularly in dense and shielded gas, the tabulated $x_{\rm H_2}$ values should be interpreted as the molecular fractions predicted by CLOUDY under the assumptions adopted in this work, rather than as a complete description of the H$_2$ abundance in the interstellar medium.

\begin{figure*}
    \centering
    \includegraphics[width=\textwidth]
    {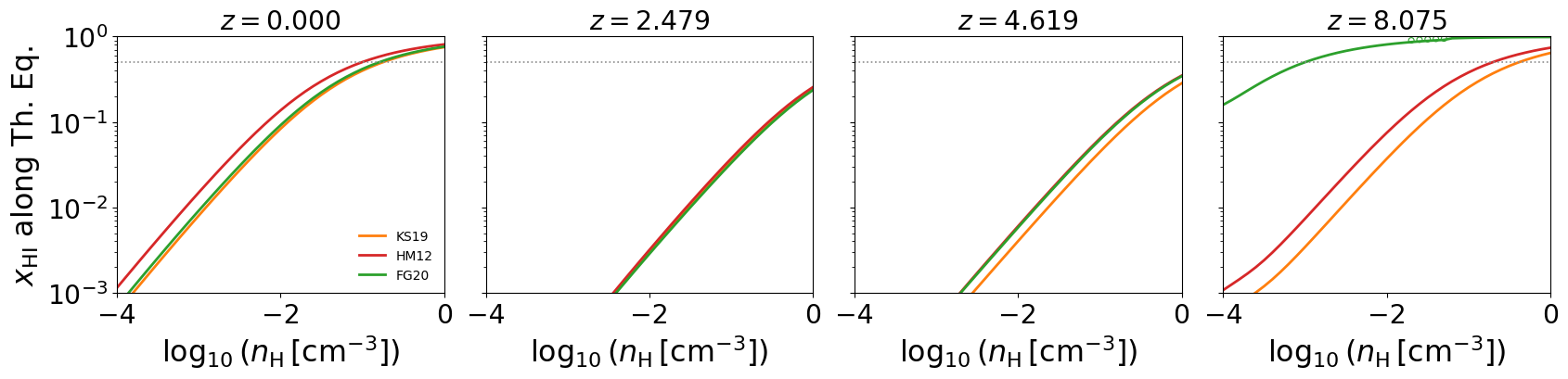}
    \caption{
    Neutral atomic hydrogen fraction for each UVB considered: KS19, HM12, and FG20. Panels correspond to the four selected redshifts. The horizontal dotted line marks $x_{\rm HI}=0.5$ and provides a reference for the transition between predominantly ionized and predominantly neutral atomic hydrogen; not every curve necessarily crosses this value within the density interval shown. The tables assume optically thin gas, so this transition should not be interpreted as a self-shielding threshold.
    }
    \label{fig:hydrogen_equilibrium}
\end{figure*}

Figure~\ref{fig:hydrogen_equilibrium} shows the resulting neutral atomic hydrogen fraction along the thermal-equilibrium tracks. This diagnostic is analogous to that used by PS20, but the comparison should be made with their unshielded UVB-only calculation, since their more complete ISM models include additional physical processes that are not considered here.

At low density, hydrogen is predominantly ionized by the metagalactic background. As the density increases, recombination becomes progressively more effective and the equilibrium neutral fraction rises. The density and
sharpness of this transition depend on the adopted UVB because the models predict different hydrogen photoionization and photoheating rates. The separation between the neutral-fraction curves therefore provides a direct physical explanation for the model-dependent equilibrium temperatures and cooling rates found in diffuse gas.

At $z=0$, the three recent UVB models predict broadly similar neutral-fraction tracks, with only a modest displacement of the transition region. The differences remain comparatively small at $z=2.479$ and $z=4.619$. 

At $z=8.075$, however, FG20 predicts a substantially larger neutral fraction over the low-density part of the equilibrium sequence, whereas the KS19 and HM12 transitions remain shifted towards higher densities. This behavior is consistent with the distinct high-redshift thermal-equilibrium structure produced by FG20.

\subsection{Element-by-element cooling}
\label{sec:results_elements}

\begin{figure*}
    \centering
    \includegraphics[width=\textwidth]
    {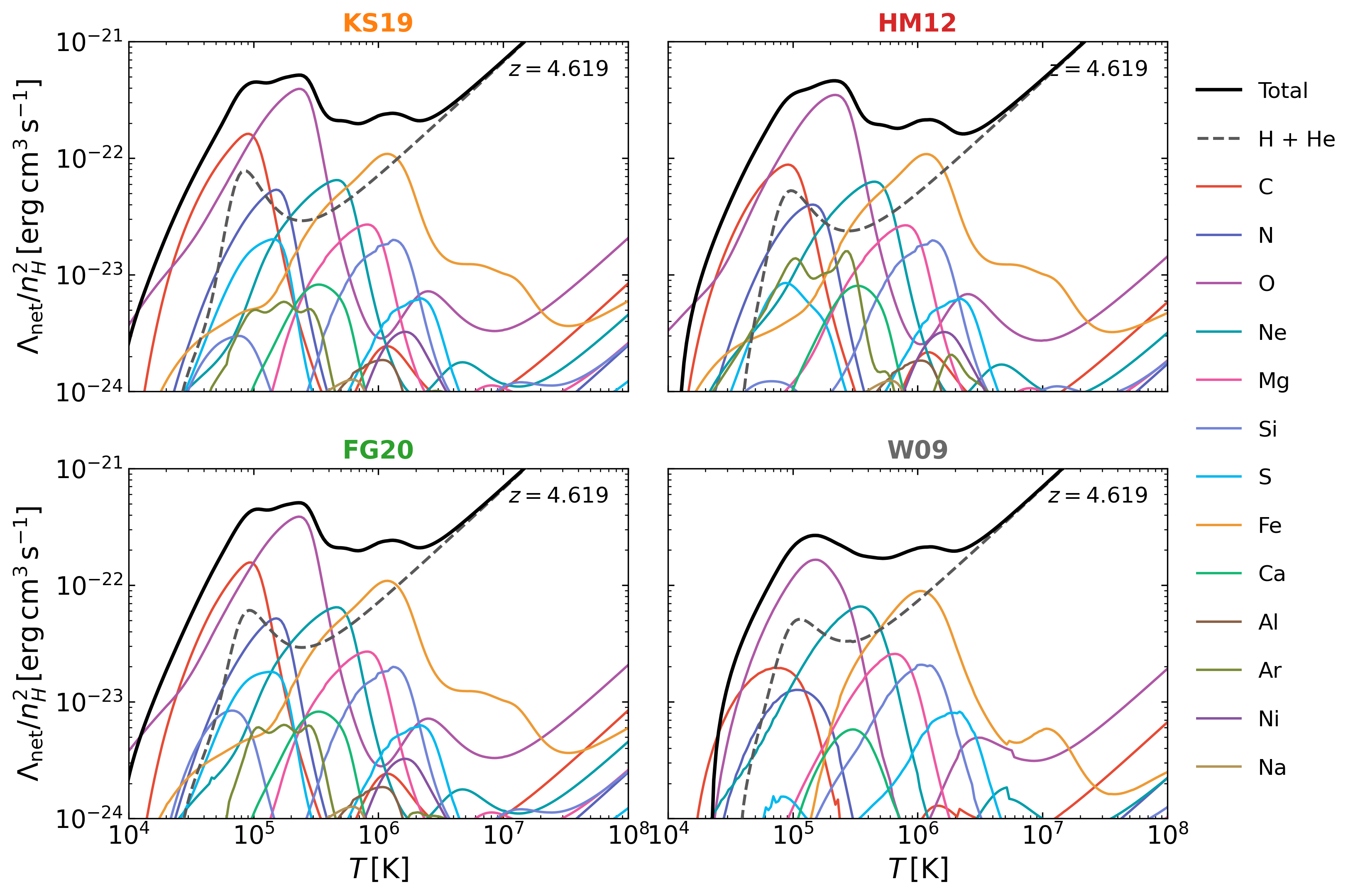}
    \caption{
    Contributions to the net cooling function at $z=4.619$ and $\log_{10}(n_{\rm H}/{\rm cm^{-3}})=-4$ for the Solar reference composition, for KS19, HM12, FG20, and \wie~ (HM01). The solid black curve shows the total net cooling function, the dashed gray curve the H and He contribution, and the colored curves the contributions of the individual heavy elements, $\Lambda_{{\rm net},i,\odot}$. The elements Na, Al, Ar, and Ni are not included in the \wie~ tables, so they do not appear in the plot.
    }
    \label{fig:elements_all}
\end{figure*}

Figure~\ref{fig:elements_all} shows the contributions of H and He and of the individual heavy elements to the net cooling function at $z=4.619$ and $\log_{10}(n_{\rm H}/{\rm cm^{-3}})=-4$, for the three new table sets and for \wie~ (HM01). We choose this redshift because it marks the onset of the large differences in the equilibrium temperature discussed in Section~\ref{sec:results_equilibrium}.

In all three new calculations, the total net cooling function is dominated by C around $T\sim10^{5}$~K, by O between $\sim10^{5}$ and $\sim3\times10^{5}$~K, by Ne around $\sim5\times10^{5}$~K, and by Fe around $\sim10^{6}$~K. The elements added in this work, Na, Al, Ar, and Ni, remain subdominant.  KS19 and FG20 are very similar, whereas HM12 shows lower C and H and He contributions around $10^{5}$~K, and its net cooling function drops more steeply at low temperature. At $T\gtrsim10^{7}$~K, the net cooling function is dominated by inverse Compton cooling off the CMB. 

The differences with respect to \wie~ are larger. In \wie, the O peak is lower by a factor of $\approx2$ and located at a lower temperature, the C contribution is lower by almost an order of magnitude, and the total net cooling function drops  at $T\approx2\times10^{4}$~K. Since the comparison with \wie~ also reflects the different CLOUDY version, atomic data, and abundance set, these differences cannot be attributed to the UVB alone. At $T\gtrsim10^{7}$~K, where Compton cooling off the CMB dominates, all four calculations converge.

\begin{figure*}
    \centering
    \includegraphics[width=0.82\textwidth]
    {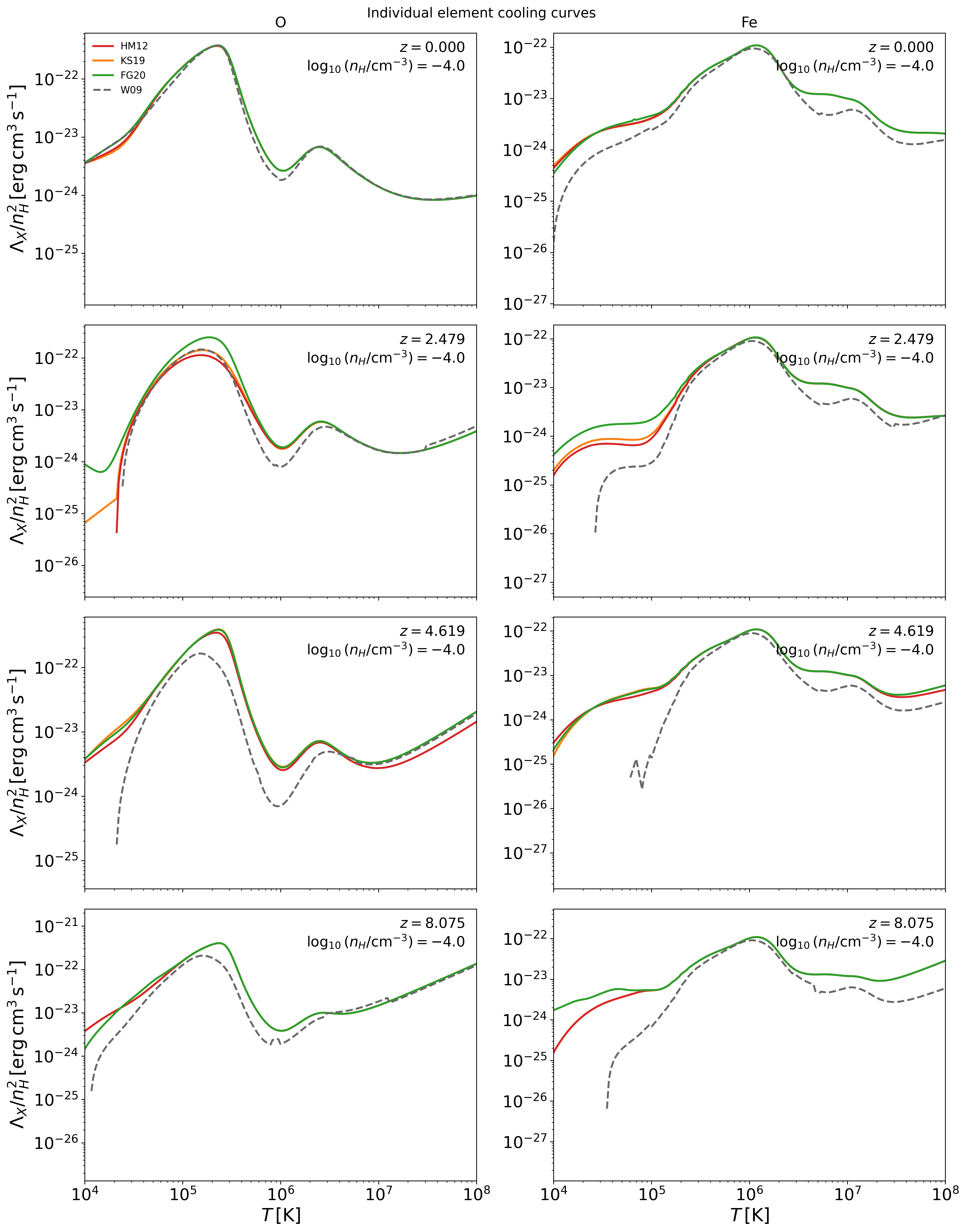}
    \caption{ Net cooling contributions of oxygen and iron, $\Lambda_{{\rm net},i,\odot}$, at $\log_{10}(n_{\rm H}/{\rm cm^{-3}})=-4$, for the four selected redshifts and UVB prescriptions. \wie~ (HM01) is shown as a gray dashed curve. At $z=8.075$, the KS19 curve lies exactly beneath the HM12 curve and is therefore not visible.
}
    \label{fig:elements_low_density}
\end{figure*}

To follow the redshift evolution of this response, Figure~\ref{fig:elements_low_density} shows oxygen and iron at the same density for the four selected redshifts. These elements contribute over broad and particularly important temperature intervals, and provide representative examples of the metal-line response to the UVB.

At the diffuse density shown in Figure ~\ref{fig:elements_low_density}, the curves separate increasingly at intermediate and high redshift. Photoionization shifts oxygen and iron towards higher ionization stages, changing the ions available for collisional excitation and therefore modifying both the amplitude and shape of their cooling functions.

Oxygen and iron are selected as physically representative examples because they are among the most important coolants: oxygen provides a substantial contribution near the characteristic metal-line cooling peak, whereas iron contributes over a broader temperature range.The UVB effect is particularly visible for iron because of its large number of ionization stages and radiative transitions. The hydrogen and helium calculations in Appendix~\ref{app:hhe_results} show that the UVB sensitivity is already present without metals; metal-line cooling modifies and can amplify this response over a broader temperature interval.

\subsection{Cooling times}
\label{sec:results_cooling_times}

Differences in the net cooling rates directly affect the time required for the gas to radiate its internal energy. Using the definition of the radiative cooling time given in Eq.~(\ref{eq:cooling_time_general}), we calculate $t_{\rm cool}$ at each point of the density-temperature grid and for each redshift using the quantities provided in our cooling tables.

%The cooling time can be defined as the ratio between the internal energy density of the gas and the radiative energy-loss rate per unit volume \citep{CimattiFEG},

%\begin{equation}
%    t_{\rm cool}
%    =
%    \frac{u_{\rm th}}
%    {\left|\mathrm{d}u_{\rm th}/\mathrm{d}t\right|}.
%    \label{eq:tcool_general}
%\end{equation}
%
%For the cooling tables used here, the net radiative energy-loss rate per unit volume is expressed as
%
%\begin{equation}
%    \left|\frac{\mathrm{d}u_{\rm th}}{\mathrm{d}t}\right|
%    =
%    n_{\rm H}^{2}\Lambda_{\rm net}(n_{\rm H},T,z),
%    \qquad \Lambda_{\rm net}>0,
%    \label{eq:net_loss_rate}
%\end{equation}

%where $\Lambda_{\rm net}$ is the net cooling coefficient. Therefore, at each point of the density-temperature grid and for each redshift, we calculate the net cooling time as

%\begin{equation}
%    t_{\rm cool}(n_{\rm H},T,z)
%    =
%    \frac{u_{\rm th}(n_{\rm H},T,z)}
%    {n_{\rm H}^{2}\Lambda_{\rm net}(n_{\rm H},T,z)}.
%    \label{eq:tcool}
%\end{equation}

For an ideal monatomic gas, we define the internal energy density in Eq.~\ref{eq:uth}. The mean particle mass is written as $\mu m_{\rm H}$, where the dimensionless quantity $\mu$ is taken directly from the corresponding Solar cooling table at each $(n_{\rm H},T,z)$ grid point. Since the gas mass density can be expressed both as $\rho_{\rm gas}=m_{\rm H}n_{\rm H}/X_{\rm H}$ and as $\rho_{\rm gas}=n_{\rm tot}\mu m_{\rm H}$, the total particle number density is

\begin{equation}
    n_{\rm tot}(n_{\rm H},T,z)
    =
    \frac{n_{\rm H}}
    {X_{\rm H}\mu(n_{\rm H},T,z)},
    \label{eq:ntot}
\end{equation}

where $X_{\rm H}$ is the hydrogen mass fraction. For the Solar reference composition, we adopt $X_{\rm H}=0.7374839$, consistent with the GASS10 abundance set.

Substituting Eq.~(\ref{eq:ntot}) into Eq.~(\ref{eq:uth}) and using the definition of the cooling time in
Eq.~(\ref{eq:cooling_time_general}), we obtain 

\begin{equation}
    t_{\rm cool}(n_{\rm H},T,z)
    =
    \frac{3k_{\rm B}T}
    {2X_{\rm H}\,
     \mu(n_{\rm H},T,z)\,
     n_{\rm H}\,
     \Lambda_{\rm net}(n_{\rm H},T,z)}.
    \label{eq:tcool_explicit}
\end{equation}

\begin{figure*}
    \centering
    \includegraphics[width=\textwidth]
    {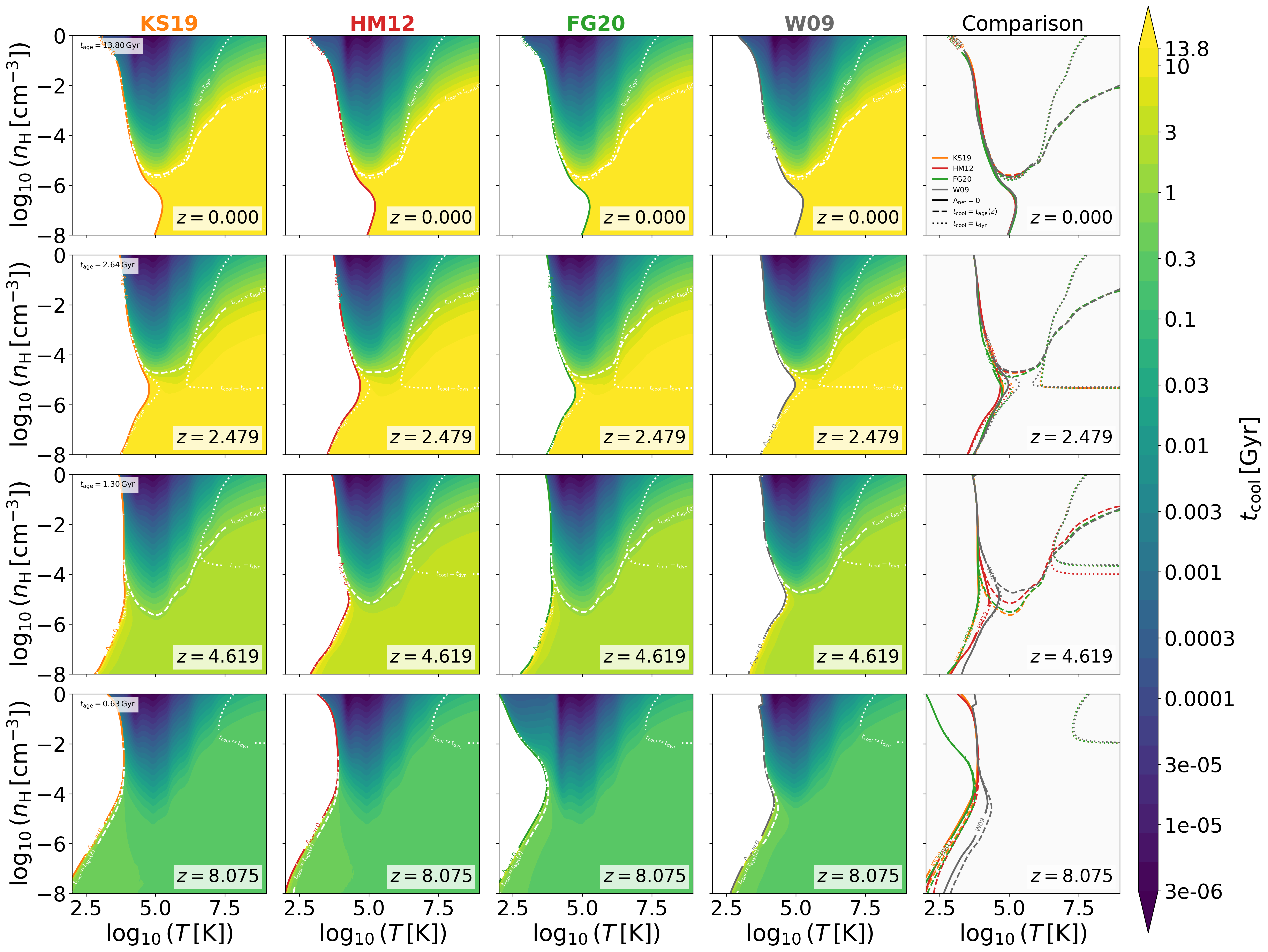}
    \caption{
    Net cooling time in the density-temperature plane for Solar composition at $z=0.000$, $2.479$, $4.619$, and $8.075$. Columns show KS19, HM12, FG20, \wie~ (HM01), and an overlay of the contours. Colors show $t_{\rm cool}$ in Gyr; white regions correspond to net heating or thermal equilibrium. Solid curves mark $\Lambda_{\rm net}=0$, dashed curves indicate where the cooling time equals the age of the Universe at the corresponding redshift $t_{\rm cool}=t_{\rm age}(z)$, and dotted curves indicate where the cooling time equals the characteristic dynamical time $t_{\rm cool}=t_{\rm dyn}$.
    }
    \label{fig:tcool_solar}
\end{figure*}

In Figure~\ref{fig:tcool_solar}, regions where $\Lambda_{\rm net}(n_{\rm H},T,z)\leq 0$, which correspond to net heating or thermal equilibrium, are left blank.

The solid curves identify thermal equilibrium. The dashed curves are obtained by solving

\begin{equation}
    t_{\rm cool}(n_{\rm H},T,z)=t_{\rm age}(z),
\end{equation}

We compute $t_{\rm age}$ assuming a flat $\Lambda$CDM cosmology with $H_0=67.74~{\rm km\,s^{-1}\,Mpc^{-1}}$, $\Omega_{\rm m}=0.309$, and $\Omega_{\Lambda}=0.691$ \citep{Planck2015}, where $H_0$ is the Hubble constant and $\Omega_{\rm m}$ and $\Omega_{\Lambda}$ are the present-day matter and dark-energy density parameters, respectively. The ages at the four selected redshifts are approximately $13.8$, $2.6$, $1.3$, and $0.63$ Gyr. The contours distinguish gas that can cool within the cosmic
time available at the corresponding redshift from gas whose radiative cooling
time is longer than that interval.

Following the classical cooling arguments of \citet{rees_cooling_1977} and \citet{white_core_1978}, we also compare the radiative cooling time with a characteristic dynamical timescale of the halo. This comparison provides a physically motivated diagnostic of how UVB-dependent shifts of the thermal-equilibrium loci in the density-temperature plane translate into changes in the region where radiative cooling is faster than the characteristic dynamical evolution of the halo, especially at high redshift.

In this framework, the thermal evolution of gas within a gravitational potential depends on the relative importance of radiative cooling and
the characteristic dynamical timescale of the halo. Following the formulation summarized by \citet{CimattiFEG}, we define the characteristic halo dynamical time as

\begin{equation}
    t_{\rm dyn}
    =
    \frac{1}{\sqrt{G\langle\rho\rangle}},
    \label{eq:tdyn}
\end{equation}

where $\langle\rho\rangle$ is the characteristic mean total mass density associated with the halo and $G$ is the gravitational constant. To express this timescale in terms of the hydrogen number density, we assume that the gas mass density represents a fixed fraction $f_{\rm gas}$ of the total mass density,

\begin{equation}
    \rho_{\rm gas}
    =
    f_{\rm gas}\langle\rho\rangle.
    \label{eq:fgas}
\end{equation}

Using the relation between gas mass density and hydrogen number density,

\begin{equation}
    \rho_{\rm gas}
    =
    \frac{m_{\rm H}n_{\rm H}}{X_{\rm H}},
\end{equation}

we obtain

\begin{equation}
    \langle\rho\rangle
    =
    \frac{m_{\rm H}n_{\rm H}}
    {X_{\rm H}f_{\rm gas}},
\end{equation}

and therefore

\begin{equation}
    t_{\rm dyn}(n_{\rm H})
    =
    \left(
    \frac{X_{\rm H}f_{\rm gas}}
    {Gm_{\rm H}n_{\rm H}}
    \right)^{1/2}.
    \label{eq:tdyn_nH}
\end{equation}

We adopt $f_{\rm gas}=0.16$, following the reference value used by \citet{CimattiFEG}, corresponding to the cosmic baryon fraction. The dotted curves in Figure ~\ref{fig:tcool_solar} satisfy $t_{\rm cool}=t_{\rm dyn}$. Regions where $t_{\rm cool}<t_{\rm dyn}$ correspond to gas that can radiate its thermal energy on a timescale shorter than the characteristic dynamical timescale of the halo. Conversely, where $t_{\rm cool}>t_{\rm dyn}$, radiative cooling proceeds more slowly than the halo dynamical timescale.

To quantify the differences in cooling time relative to \wie, we define

\begin{equation}
    \Delta\log_{10}t_{\rm cool}
    =
    \log_{10}\left(
    \frac{t_{\rm cool,model}}{t_{\rm cool,W09}}
    \right).
\end{equation}

Here, $t_{\rm cool,model}$ is the cooling time computed for a given UVB model, while $t_{\rm cool,W09}$ is the corresponding cooling time from the \wie~ tables. Positive values of $\Delta\log_{10}t_{\rm cool}$ indicate longer cooling times than in \wie, while negative values indicate shorter cooling times.

\begin{figure*}
    \centering
    \includegraphics[width=0.82\textwidth]
    {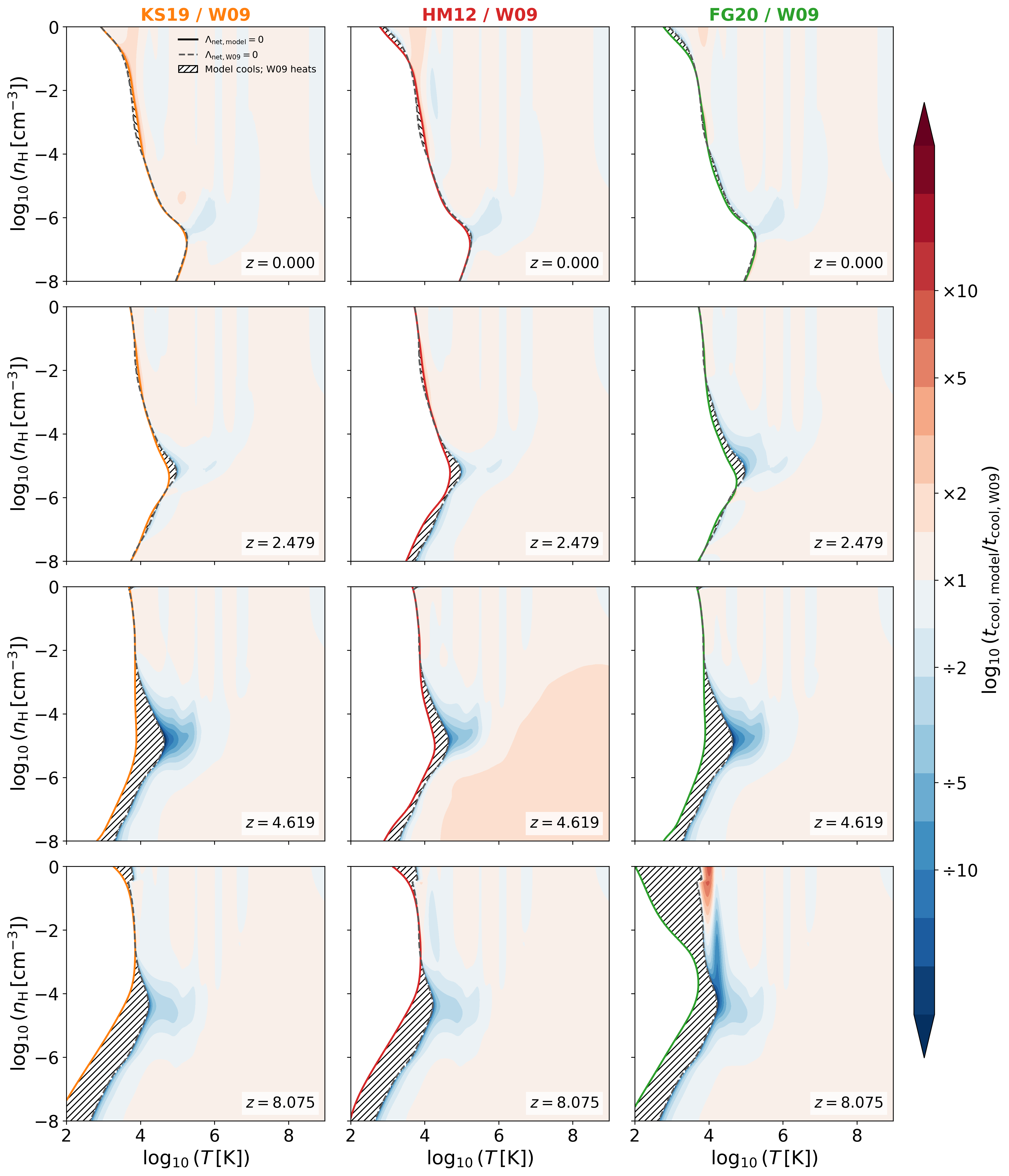}
    \caption{
    Cooling-time differences between KS19, HM12, and FG20 and \wie~ (HM01)
    for Solar composition. Colors show
    $\log_{10}(t_{\rm cool,model}/t_{\rm cool,W09})$ where both models
    predict net cooling. Solid and dashed curves mark the model and \wie~
    equilibrium boundaries. Hatched regions indicate disagreement on the sign
    of the net rate; gray regions are undefined or masked.
    }
    \label{fig:tcool_ratio_solar}
\end{figure*}

For the Solar-composition comparison in Figure ~\ref{fig:tcool_ratio_solar}, at $z=0$ the cooling-time differences are generally small: over the diffuse-density interval, half of the grid points show $\Delta\log_{10}t_{\rm cool} =
\log_{10}(t_{\rm cool,model}/t_{\rm cool,W09})$ between $-0.01$ and $+0.03$~dex, and the models disagree on the sign of the net rate over less than $1\%$ of that region. In a limited region close to the thermal equilibrium boundary, however, the new models predict shorter cooling times than \wie, by up to a factor of $\approx 2$. This region occupies $-7 \lesssim \log_{10}(n_{\rm H}/{\rm cm^{-3}}) \lesssim -4$ and $4.1 \lesssim \log_{10}(T/{\rm K}) \lesssim 5.8$ at $z=0$, i.e.\ the diffuse, photoionized gas typical of the intergalactic medium, only its warmer edge overlapping the warm-hot intergalactic medium (WHIM) regime, $T \sim 10^5$-$10^7$~K.

This behavior becomes markedly stronger with redshift, while shifting to progressively lower temperatures: the affected region spans $-7.8 \lesssim \log_{10}(n_{\rm H}/{\rm cm^{-3}}) \lesssim -3.5$, $3.9 \lesssim \log_{10}(T/{\rm K}) \lesssim 5.3$ at $z\simeq 2.5$; $-7.4 \lesssim \log_{10}(n_{\rm H}/{\rm cm^{-3}}) \lesssim -3.6$, $3.6 \lesssim \log_{10}(T/{\rm K}) \lesssim 5.4$ at $z\simeq 4.6$; and $-7.6 \lesssim \log_{10}(n_{\rm H}/{\rm cm^{-3}}) \lesssim -3.2$, $2.9 \lesssim \log_{10}(T/{\rm K}) \lesssim 4.9$ at $z\simeq 8.1$. Within this region, cooling times fall short of \wie~ by up to a factor of $\approx 10$-$11$ already at $z\simeq 4.6$, remaining at a similar level at $z\simeq 8.1$. At $z\gtrsim 4.6$ this region of parameter space lies entirely below the WHIM temperature range, corresponding to the diffuse IGM rather than to shock-heated gas. At $z\simeq 4.6$, HM12 also departs markedly from KS19 and FG20 in its median offset, which reaches $+0.18$~dex (a factor of $\approx 1.5$ in $t_{\rm cool}$) over the diffuse-density interval, consistent with the distinct behavior of HM12 already noted in Section~\ref{sec:results_equilibrium}. The fraction of the density-temperature plane where the models disagree on the sign of the net rate grows steadily with redshift, from $0.4$-$0.7\%$ at $z=0$ to $8.8$-$9.9\%$ at $z\simeq 8.1$ within the diffuse-density interval.

% SUMMARY AND DISCUSSION

\section{Conclusion and outlook}
\label{sec:summary}

We have presented new cooling and heating tables for optically thin gas in photoionization equilibrium exposed to the HM12, KS19, and FG20 redshift-dependent UVB radiation models. The tables cover the redshift range available for each individual UVB, with the sampling described in Section~\ref{sec:grid}, hydrogen number densities $-8 \leq \log_{10}(n_{\rm H}/{\rm cm^{-3}}) \leq 0$, and temperatures $2 \leq \log_{10}(T/{\rm K}) \leq 8.982$. They retain separate cooling, heating, and net rates, tabulate H and He over an extended helium abundance grid, provide HI, HII, and H$_2$ fractions, and include element-by-element contributions for 13 chemical elements: Aluminium, Argon, Calcium, Carbon, Iron, Magnesium, Neon, Nickel, Nitrogen, Oxygen, Silicon, Sodium, Sulphur. The data are stored in the HDF5 format designed to remain compatible with the structure of the \wie~ tables.

The main results can be summarized as follows.

\begin{enumerate}
    \item At $z=0$, KS19, HM12, FG20, and \wie~ predict a similar overall heating-cooling structure. The differences are concentrated in diffuse gas and near thermal equilibrium, whereas the models converge at high temperature and high density.

    \item The model dependence grows towards high redshift. The differences become visible at $z\simeq2.5$, increase at $z\simeq4.6$, and are largest at $z\simeq8.1$, where the UVB emissivity and reionization histories are the most uncertain.

    \item The hydrogen and helium calculations collected in Appendix~\ref{app:hhe_results} recover the same qualitative UVB-dependent trends found in the Solar-composition analysis. The differences therefore do not originate exclusively from metal-line cooling, although metals modify and can amplify their magnitude and extent.

    \item FG20 UVB model produces the most distinctive high-redshift behavior.

    \item Changes in net rates propagate into the cooling time and shift the regions that satisfy $t_{\rm cool}<t_{\rm age}$ and $t_{\rm cool}<t_{\rm dyn}$ conditions. The differences from \wie~ remain small over most of the diffuse density-temperature plane, but grow near the thermal-equilibrium boundary, where the new tables predict shorter cooling times, by up to a factor of $\approx 2$ at $z=0$ and up to a factor of $\approx 10$-$11$ already at $z\simeq 4.6$, remaining at a similar level at $z\simeq 8.1$. At these higher redshifts, this region lies below the WHIM temperature range, in the diffuse IGM rather than in shock-heated gas. The models also disagree on the sign of the net rate over a growing fraction of the diffuse plane, from below $1\%$ at $z=0$ to nearly $10\%$ at $z\simeq 8.1$.

\end{enumerate}

The tables assume a spatially uniform UVB, optically thin gas, and ionization equilibrium. They do not include self-shielding, local stellar or AGN sources, dust attenuation, cosmic rays, or non-equilibrium chemistry. Their direct application to dense interstellar gas and to regions undergoing reionization should therefore be treated with caution. Within this domain, the release provides an updated implementation of the element-by-element cooling framework introduced by \wie, together with a consistent set of calculations for several modern metagalactic UV/X-ray background models. The common numerical setup of the HM12, KS19, and FG20 tables also enables the sensitivity to the adopted UVB prescription to be assessed directly.

The deliberately simplified physical setup adopted here is intended to isolate the impact of the metagalactic UV/X-ray background,  where keeping the record of the heating and cooling contributions separately retains the flexibility to combine the cooling tables with additional physical modules in cosmological simulations. In applications where the time-dependent chemical evolution of the gas becomes important, the tables can be complemented by a non-equilibrium chemical network. 

One example is the framework developed by \citet{maio07, maio10}, which follows the time-dependent evolution of atomic and molecular species through a network of chemical reactions coupled to the thermal evolution of the gas. Such a treatment would complement the equilibrium ion and molecular fractions tabulated here and provide a more appropriate description of the chemical evolution of the gas in regimes where the assumption of chemical equilibrium becomes inadequate. Additional cooling processes can also be included separately. For example, the dust model described by \citet{Granato2021} includes the cooling of hot gas through collisional energy transfer from the gas to dust grains, followed by infrared emission from the grains.  In addition, extending the covered helium mass fraction compared to previous works makes these tables ready to be used in galaxy formation simulations which include more extreme regions within ISM and IGM, as for example cores of Seyfert galaxies, where high helium mass fractions are observed \citep{2022MNRAS.514.5506D}. All these examples illustrate how the present UVB-dependent cooling tables can serve as one component of a broader physical model, in which processes not included in the tables themselves are treated by dedicated modules.

% CONCLUSION

%We have presented an updated set of element-by-element radiative cooling and photoheating tables for cosmological simulations, following the methodology introduced by \wie. The new calculations extend this framework to modern UV/X-ray background prescriptions and a recent version of CLOUDY, while providing separate cooling and heating rates, an expanded helium grid, hydrogen ionization fractions, and additional elemental contributions.

%The comparison with \wie~ quantifies how the predicted thermal properties change when the underlying table calculations are updated. The differences are most pronounced in diffuse, photoionized gas and at high redshift, and can propagate into the thermal-equilibrium temperature and net cooling
%timescale. In some regions of parameter space, the updated calculations also predict a different balance between net cooling and net heating.
%Comparisons among HM12, KS19, and FG20 further quantify the sensitivity of the tables to the adopted metagalactic UV/X-ray background within a common numerical framework. 

% ACKNOWLEDGMENT

\section*{Acknowledgments}
LDF thanks Aura Obreja for her precious help during the initial stages of this work. LDF and GSK acknowledge support from the Scalable Parallel Astrophysical Codes for Exascale (SPACE) Centre of Excellence, which received funding from the European High Performance Computing Joint Undertaking (JU) and Belgium, Czech Republic, France, Germany, Greece, Italy, Norway, and Spain under Grant Agreement No. 101093441.
MV acknowledges support from the Fondazione ICSC National Recovery and Resilience Plan (PNRR) Project ID CN-00000013 "Italian Research Center on High-Performance Computing, Big Data and Quantum Computing" funded by MUR Missione 4 Componente 2 Investimento 1.4: "Potenziamento strutture di ricerca e creazione di campioni nazionali di R\&S (M4C2-19 )" - Next Generation EU (NGEU). MV acknowledges partial financial support from the INFN Indark Grant and the "Astrofisica Fondamentale 2024" INAF Grant "Black Hole Dynamics and Galaxy Formation from Cosmological Simulations". 
IK was supported by the Simons Foundation via the Simons Investigator Award to A. A. Schekochihin.
%DATA AVAILABILITY

\section*{Data Availability}
\label{sec:data_availability}
The complete HM12, KS19, and FG20 table sets are publicly available at \url{https://homepages.usm.uni-muenchen.de/OpenGadget/CoolingTables/}.

% APPENDIX

\appendix
\section{Detailed comparison of the UVB models}
\label{app:uvb_models}

Here we describe in detail how the UVB models used in this work differ from one another.

\subsection{Stellar population synthesis}

The models adopt different stellar population synthesis prescriptions to describe the radiation emitted by star-forming galaxies. HM01 uses a Bruzual \& Charlot stellar spectrum with a Salpeter initial mass function (IMF; \citealt{salpeter_luminosity_1955}), a fixed metallicity of $Z=0.2\,Z_\odot$, and a stellar age of $0.5$~Gyr \citep{haardt_modelling_2001}. This corresponds to a single representative stellar population.

HM12 also uses the Bruzual \& Charlot GALAXEV library \citep{bruzual_stellar_2003} and assumes a Salpeter IMF over the mass range $0.1$-$100\,M_\odot$. In this case, however, the galaxy spectrum is not associated with a single stellar age. Instead, it is obtained by convolving the evolution of simple stellar populations with the cosmic star-formation history. HM12 also includes a redshift-dependent stellar metallicity,
\begin{equation}
    Z(z)=Z_\odot\,10^{-0.15z},
\end{equation}
so that the stellar populations become progressively more metal-poor at high redshift.

FG20 instead uses BPASS v2.2.1 \citep{eldridge_binary_2017}, which includes interacting binary stars, assuming continuous star formation over 300\,Myr and a fixed stellar metallicity $Z = 0.1\,Z_\odot$. The adopted IMF is a broken power law with index 1.30 below 0.5\,$M_\odot$ and 2.45 above, up to 300\,$M_\odot$. Including interacting binaries keeps hot, massive stars active for longer than in single-star models, which extends the production of ionizing photons after a burst \citep{eldridge_binary_2017}.

KS19 instead uses the single-star \textsc{Starburst99} synthesis model \citep{1999Leitherer}. It assumes a Salpeter IMF between $0.1$ and $100\,M_\odot$ and a fixed metallicity of $Z=0.008$, corresponding to approximately $0.4\,Z_\odot$. As in HM12, the cosmic galaxy emissivity is obtained by convolving the stellar spectra with the redshift-dependent star-formation rate density.

\subsection{Galaxy emissivity and star-formation history}

HM01 do not reconstruct the cosmic star-formation rate density independently. They adopt the star-formation history of \citet{madau_high-redshift_1996} to normalize the stellar emission at 1500\,\AA, and the ionizing emissivity then follows from the assumed stellar template and escape fraction.

To build the SFRD, HM12 combine far-UV luminosity densities from three sets of surveys: \citet{schiminovich_galex-vvds_2005} for $0 \le z \le 2$, \citet{reddy_steep_2009} at $z \approx 2.3$--3, and \citet{bouwens_ultraviolet_2011} between $z = 3.8$ and 8. After dust correction, these densities are matched by iterating on the SFRD until the far-UV output of the stellar population model agrees with them. The resulting fit is
\begin{equation}
{\rm SFRD}(z) = \frac{6.9\times10^{-3} + 0.14\,(z/2.2)^{1.5}}{1 + (z/2.7)^{4.1}}\ M_\odot\,{\rm yr^{-1}\,Mpc^{-3}}.
\label{eq:sfrd_hm12}
\end{equation}

FG20 does not normalize the galaxy term through an explicit SFRD fit. Instead, a tabulated comoving emissivity at 1500\,\AA\ is interpolated in redshift: up to $z = 2$ it is taken from \citet{chiang_broadband_2019}, and from $z = 2.1$ onwards it follows from integrating observed rest-UV luminosity functions to $M_{\rm UV} = -13$. Since FG20 uses luminosity densities before any dust correction, the galaxy contribution is anchored to the UV radiation that actually leaves galaxies.

In KS19, the SFRD and the dust attenuation are determined together by fitting galaxy luminosity functions measured at several wavelengths. This reduces the degeneracy between the intrinsic star-formation rate and the fraction of UV radiation absorbed by dust. The adopted fit is
\begin{equation}
{\rm SFRD}(z) = 10^{-2}\,\frac{(2.01 + 8.48\,z)}{1 + (z/2.5)^{3.09}}\ M_\odot\,{\rm yr^{-1}\,Mpc^{-3}}.
\label{eq:sfrd_ks19}
\end{equation}

\subsection{Dust attenuation}

HM01 do not include an explicit dust attenuation curve in the ionizing galaxy emissivity. They define $f_{\rm esc}$ through the 912-to-1500\,\AA\ luminosity ratio, comparing its escaping and intrinsic values,
\begin{equation}
f_{\rm esc} = \frac{[L_{912}/L_{1500}]_{\rm esc}}{[L_{912}/L_{1500}]_{\rm int}}.
\label{eq:fesc_hm01}
\end{equation}
Only this ratio enters, so $f_{\rm esc}$ quantifies the additional suppression of Lyman-continuum photons relative to the non-ionizing UV, independently of the overall normalization. Any attenuation acting equally at both wavelengths therefore cancels out.

HM12 uses the \citet{calzetti_dust_2000} attenuation law. The attenuation at $1500$~\AA\ is set to $A_{\mathrm{FUV}}=1$~mag for $z\leq2$, while at higher redshift it is described by
\begin{equation}
    A_{\mathrm{FUV}}(z)=2.5\log_{10}\left(1+\frac{1.5}{z-1}\right).
\end{equation}
This prescription produces a decreasing dust correction toward high redshift.

FG20 adopts the attenuation curve of \citet{reddy_mosdef_2015}, calibrated using observations of Lyman-break galaxies at $z \sim 3$. It assumes a fixed color excess of $E(B-V) = 0.129$. In its holes model, dust attenuates the radiation along obscured sightlines, whereas photons escaping through clear channels are assumed to be unaffected by dust.

KS19 uses the LMC2 extinction curve of \citet{gordon_quantitative_2003}. Its redshift-dependent far-UV attenuation is
\begin{equation}
    A_{\mathrm{FUV}}(z)=
    \frac{1.42+0.93z}{1+(z/2.08)^{2.2}}.
\end{equation}
This attenuation history is determined consistently with the SFRD from the multi-wavelength galaxy luminosity functions.

\subsection{Escape of ionizing radiation from galaxies}

The escape fraction determines how much of the ionizing radiation produced by stars reaches the IGM. HM01 sets the escape fraction to $f_{\mathrm{esc}}=0.1$ to normalize the stellar ionizing contribution.

In HM12, the luminosity-weighted escape fraction applied to photons between 1 and 4\,Ryd grows as a power of $(1+z)$:
\begin{equation}
f_{\rm esc}(z) = 1.8\times10^{-4}\,(1+z)^{3.4}.
\label{eq:fesc_hm12}
\end{equation}
This gives $\approx 0.8\%$ at $z = 2$, $2.6\%$ at $z = 3.3$ and more than $20\%$ at $z > 7$. No stellar photons above 4\,Ryd are allowed to escape. This steep rise is what allows galaxies to sustain the H\,\textsc{i}-ionizing background at early times and to complete H\,\textsc{i} reionization.

FG20 describes the escape of ionizing photons through a holes model. A fraction of the sightlines is assumed to be free of both neutral-gas absorption and dust attenuation. The absolute escape fraction is
\begin{equation}
    f_{\mathrm{esc}}(z)=
    \min\left[0.01\left(\frac{1+z}{4}\right)^{2.5},\,0.2\right].
\end{equation}
It is therefore $1\%$ at $z=3$, increases with redshift, and is limited to a maximum value of $20\%$. Unlike the explicit prescriptions in HM12 and KS19, FG20 does not impose the same sharp cutoff of the stellar spectrum above $4$~Ryd, although the stellar contribution at these energies remains small.

KS19 adopts
\begin{equation}
    f_{\mathrm{esc}}(z)=
    \frac{10^{-5}}{6.5\times10^{-5}+\exp(-2.4z)}.
\end{equation}
The escape fraction rises rapidly between $z\sim3$ and $z\sim5.5$ and approaches an asymptotic value of about $15\%$ at $z\gtrsim6$. As in HM12, the escaping stellar emissivity is set to zero above $4$~Ryd.

\subsection{AGN emission}

The models also differ in their AGN luminosity functions and spectral energy distributions. HM01 uses the optical QSO luminosity function of \citet{boyle_2df_2000}. Below $1050$~\AA, the QSO spectrum is described by a power law with $L_\nu\propto\nu^{-1.8}$, based on the results of \citet{laor_soft_1997} and \citet{zheng_composite_1998}.

HM12 model the quasar contribution with a fit to the 1\,Ryd emissivity derived from the bolometric QSO luminosity function of \citet{hopkins_observational_2007}. The UV spectral energy distribution is a broken power law,
\begin{equation}
L_\nu \propto
\begin{cases}
\nu^{-0.44}, & \lambda > 1300\,\text{\AA},\\
\nu^{-1.57}, & \lambda < 1300\,\text{\AA},
\end{cases}
\label{eq:sed_hm12}
\end{equation}
based on \citet{vandenberk2001} and  \citet{telfer2002}. The ionizing slope is therefore harder than the fiducial $\nu^{-1.8}$ adopted in HM01 and KS19.

FG20 updates the AGN emissivity using the bolometric luminosity function of \citet{shen_bolometric_2020}. At $\lambda \ge 912$\,\AA\ the spectrum follows the mean quasar template of \citet{krawczyk_mean_2013}, while between 600 and 912\,\AA\ it is described by $L_\nu \propto \nu^{-1.7}$. Between 600\,\AA\ and 0.5\,keV, the UV and X-ray components are joined by a single power law whose normalization follows the observed optical-to-X-ray ratio $\alpha_{\rm OX}$. This choice softens the spectrum above 600\,\AA\ and yields a lower He\,\textsc{ii}-ionizing flux than in HM12 and KS19. At X-ray energies, the spectrum is an exponentially truncated power law with a Compton reflection component computed with PEXRAV \citep{magdziarz_angle-dependent_1995}.

KS19 derives the type-1 QSO emissivity from a compilation of observed QSO luminosity functions. The extreme-UV spectrum is represented by a broken power law with a free spectral index. The fiducial model adopts $L_\nu\propto\nu^{-1.8}$, but KS19 also provides models with different slopes. These variations mainly affect the He~II-ionizing part of the UVB and the thermal evolution associated with He~II reionization.

\subsection{X-ray background and obscured AGN}

HM01 also extend the calculation to X-ray energies. Type-2 spectra are obtained from Monte Carlo simulations of absorbed sources averaged over a log-normal $N_{\rm H}$ distribution, and the type-2/type-1 ratio evolves with redshift. The resulting model is compared with HEAO-1 and BeppoSAX background data and with X-ray number counts.

HM12 combines hard X-ray luminosity functions with a distribution of absorbing hydrogen columns. The spectra of obscured AGN are reprocessed through the absorbing material, and the resulting population is calibrated to reproduce measurements of the cosmic X-ray background, including those from HEAO-1 and BeppoSAX.

In FG20, the intrinsic X-ray spectrum of unobscured AGN is the cut-off power law with reflection described in Section~A.5 ($\Gamma = 1.8$, $E_{\rm c} = 300$\,keV). The spectrum averaged over the whole AGN population follows the parametric form of \citet{sazonov_quasars_2004}, with $E_1 = 67$\,keV, and the template for obscured sources is obtained as the difference between the two. An obscured fraction $f_{\rm obsc} = 0.75$ reproduces the relative intensities of the UV and X-ray backgrounds.

KS19 includes a type-2 QSO component and a blazar component in addition to type-1 QSOs. The adopted spectrum is a modified version of the \citet{sazonov_quasars_2004} template and extends from X-ray to TeV energies. Its normalization is constrained using measurements from Chandra, INTEGRAL, RXTE, Swift-BAT, and Fermi/LAT. KS19 therefore covers a wider energy interval than the other models, although these high-energy photons have a limited direct effect on the ionization of low-density gas.

\subsection{Intergalactic absorption}

The propagation of ionizing radiation depends on the distribution of intergalactic absorbers. HM01 approximates the H~I column-density distribution with a single power-law dependence, $f(N_{\mathrm{HI}})\propto N_{\mathrm{HI}}^{-1.5}$, combined with an independently specified redshift evolution. This provides a simple description of the effective intergalactic opacity.

HM12 describe the absorbers with a piecewise power law in $N_{\rm HI}$ and $(1+z)$, using five classes: low- and high-column Ly$\alpha$ forest, Lyman-limit systems, super Lyman-limit systems and damped Ly$\alpha$ systems. The redshift scaling changes at $z = 1.56$ for all classes and, for the Ly$\alpha$ forest, again at $z = 5.5$. The parameters are constrained by absorber statistics and by the mean free path of 1\,Ryd photons measured by \citet{prochaska2009}.

FG20 uses a modified version of the distribution adopted by \citet{puchwein_consistent_2019}, which is itself based on the HM12 form. The main modification is applied at $z\leq1.56$, where the redshift evolution is changed from $f(N_{\mathrm{HI}},z)\propto(1+z)^{0.16}$ to $f(N_{\mathrm{HI}},z)\propto(1+z)$. This stronger evolution provides a better match to low-redshift measurements of Lyman-limit systems and changes the mean free path of ionizing photons.

KS19 adopts the analytical fitting function of \citet{inoue_updated_2014}. The same prescription is used over the full redshift range of the model, including its extrapolation beyond $z=6$. Differences in the column-density distributions modify the mean free path of ionizing photons and therefore the amplitude and shape of the transmitted UVB.

\subsection{Reionization histories}

The HM01 model was not calibrated to produce a specific H~I or He~II reionization history. By contrast, HM12 presents a minimal reionization model in which H~I and He~II reionization are completed at $z=6.7$ and $z=2.8$, respectively. The corresponding electron-scattering optical depth is $\tau_{\mathrm{e}}=0.084$, consistent with the WMAP7 constraints available at the time.

The fiducial FG20 effective reionization model adopts $z_{\mathrm{rei,HI}}=7.8$ and $z_{\mathrm{rei,HeII}}=3.5$, with $\tau_{\mathrm{e}}=0.054$, consistent with Planck 2018. FG20 also provides effective photoionization and photoheating rates designed for simulations that assume a spatially uniform UVB and ionization equilibrium. These effective rates impose the desired mean reionization history and heating even though real reionization is spatially inhomogeneous.

In the fiducial KS19 model, H~I reionization is completed at $z=5.8$, while He~II reionization is completed at $z=2.8$. The resulting electron-scattering optical depth is approximately $\tau_{\mathrm{e}}=0.055$, consistent with the Planck constraints used in that work.

\subsection{Comparison of the spectral shapes}

\begin{figure*}[t]
    \centering
    \includegraphics[width=\textwidth]{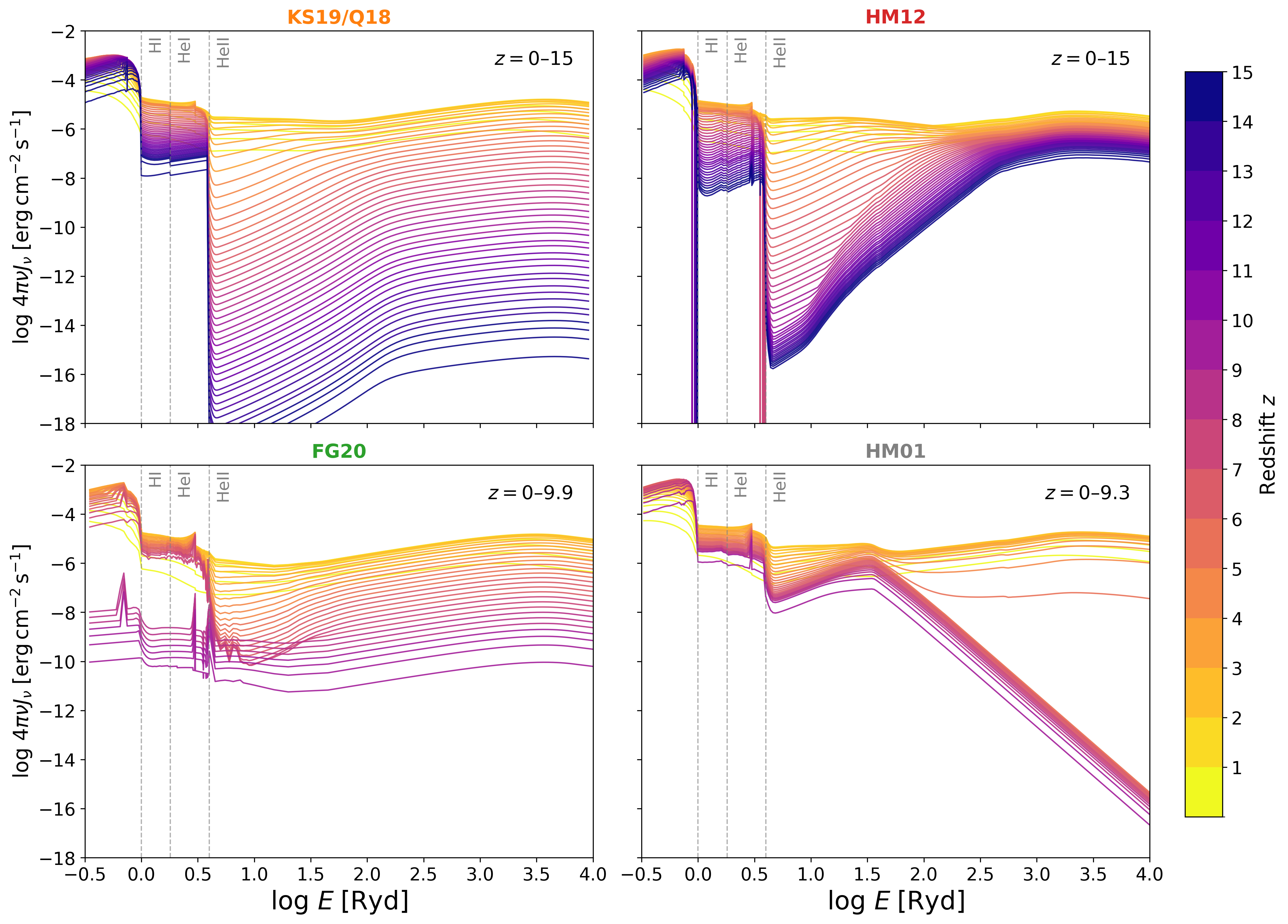}
    \caption{
        Redshift evolution of the spectral energy distributions of the four UVB models considered in this work: KS19/Q18 (top left), HM12 (top right), FG20 (bottom left), and HM01 (bottom right). The spectra are shown as a function of photon energy, expressed in Rydberg units, with colors indicating redshift according to the color bar. Vertical dashed lines mark the ionization thresholds of H~I (1 Ryd), He~I (1.81 Ryd), and He~II (4 Ryd). The models cover different redshift intervals, as indicated in each panel.
    }
    \label{fig:UVB_all_redshifts}
\end{figure*}

To illustrate the redshift evolution of the different UVB prescriptions,  Figure~\ref{fig:UVB_all_redshifts} shows their spectral energy distributions over the full redshift range available for each model. The spectra exhibit substantial variations in both intensity and shape as a function of redshift, particularly at energies above the He\,\textsc{ii} ionization threshold. These differences reflect the distinct assumptions regarding the evolution of galaxy and AGN emissivities, the escape of ionizing radiation, and intergalactic absorption. 

While Figure~\ref{fig:UVB_all_redshifts} illustrates the redshift evolution of each UVB prescription individually, Figure~\ref{fig:UVB_6panel_paper} provides a direct comparison of the different models at selected redshifts. The models are relatively similar at low redshift, although differences remain around the H\,\textsc{i}, He\,\textsc{i}, and He\,\textsc{ii} ionization thresholds. At $z=0$ and $z=3$, AGN provide an important fraction of the ionizing background and the predicted spectra are therefore broadly comparable. The differences increase at $z\gtrsim5$, when the evolution of the galaxy emissivity and escape fraction becomes more important. In particular, the models predict different intensities between $1$ and $4$~Ryd, which directly affect the ionization of H~I, He~I, and many metal ions.

At $z=5$ and $z=8$, the different decline beyond the He~II edge also reflects the assumed stellar spectra and the treatment of radiation above $4$~Ryd. The HM01 total spectrum can be separated into a softer galaxy component and a harder quasar component. FG20 shows the effect of the BPASS binary-star spectrum, while HM12 and KS19 display the sharp suppression of the escaping stellar contribution above $4$~Ryd. At still higher redshift, the differences between HM12 and KS19 become large because of their different galaxy emissivities, escape fractions, and reionization histories. These spectral differences are particularly relevant for cooling calculations: even when two models have a similar intensity near the H~I edge, a different spectral slope at higher energies can produce different ion fractions and photoheating rates for helium and metals.

Overall, HM01 provides the simplest and oldest description among the models considered here. HM12 introduces evolving stellar populations, a reconstructed SFRD, and a more detailed treatment of intergalactic and AGN absorption. FG20 uses harder binary-star spectra, a directly constrained UV emissivity, and updated AGN and absorber populations. KS19 combines a multi-wavelength reconstruction of the galaxy emissivity with a broad AGN spectrum extending to gamma-ray energies. These differences are important for net cooling calculations because they change the abundance of photons above the ionization thresholds of H, He, and metal ions. As a result, each model can produce different ionization fractions, equilibrium temperatures, and net cooling rates at the same density, temperature, metallicity, and redshift.

\section{Hydrogen and helium calculations}
\label{app:hhe_results}

The main analysis adopts a fixed Solar reference composition at all redshifts in order to compare the UVB prescriptions using a homogeneous chemical composition. In this appendix, we collect the corresponding calculations including only hydrogen and helium, adopting a helium mass fraction $Y=0.248$. Since the composition contains only hydrogen and helium, the corresponding hydrogen mass fraction is $X_{\rm H}=1-Y=0.752$. The purpose of this comparison is to test whether the trends identified in the main analysis depend on the presence of metal-line cooling.

\subsection{Thermal-equilibrium comparison}
\label{app:hhe_equilibrium}

We first compare the hydrogen and helium thermal-equilibrium loci of KS19, HM12, and FG20, restricting the quantitative comparison to $-8\leq\log_{10}(n_{\rm H}/{\rm cm}^{-3})\leq-3$. When more than one zero crossing is present at the same density, the highest-temperature crossing is used only for the purpose of constructing a single comparison branch.

At $z=0$, the three UVB prescriptions give very similar hydrogen and helium equilibrium temperatures: the median pairwise differences are $0.012$-$0.038$ dex and the maximum pairwise difference remains below $0.07$ dex. At $z=2.479$, the median differences remain modest, $0.014$-$0.051$ dex, although larger local deviations occur at the lowest densities.

The strongest separation in the hydrogen and helium comparison occurs at $z=4.619$. HM12 predicts systematically higher equilibrium temperatures than KS19 and FG20, with median offsets of approximately $0.15$ and $0.28$ dex, respectively. The HM12-FG20 difference corresponds to a factor of approximately $1.9$ in $T_{\rm eq}$, with a maximum offset of about $0.30$ dex. At $z=8.075$, the median pairwise differences decrease again to approximately $0.02$-$0.07$ dex, although local differences can reach approximately $0.2$ dex.

These results demonstrate that model-to-model differences are already present when only H and He are included. The qualitative sensitivity to the adopted UVB therefore does not originate exclusively from metal-line cooling, although metals can modify and amplify the differences seen in the Solar reference calculations.

\subsection{Cooling times}
\label{app:hhe_cooling_times}

\begin{figure*}
    \centering
    \includegraphics[width=\textwidth]
    {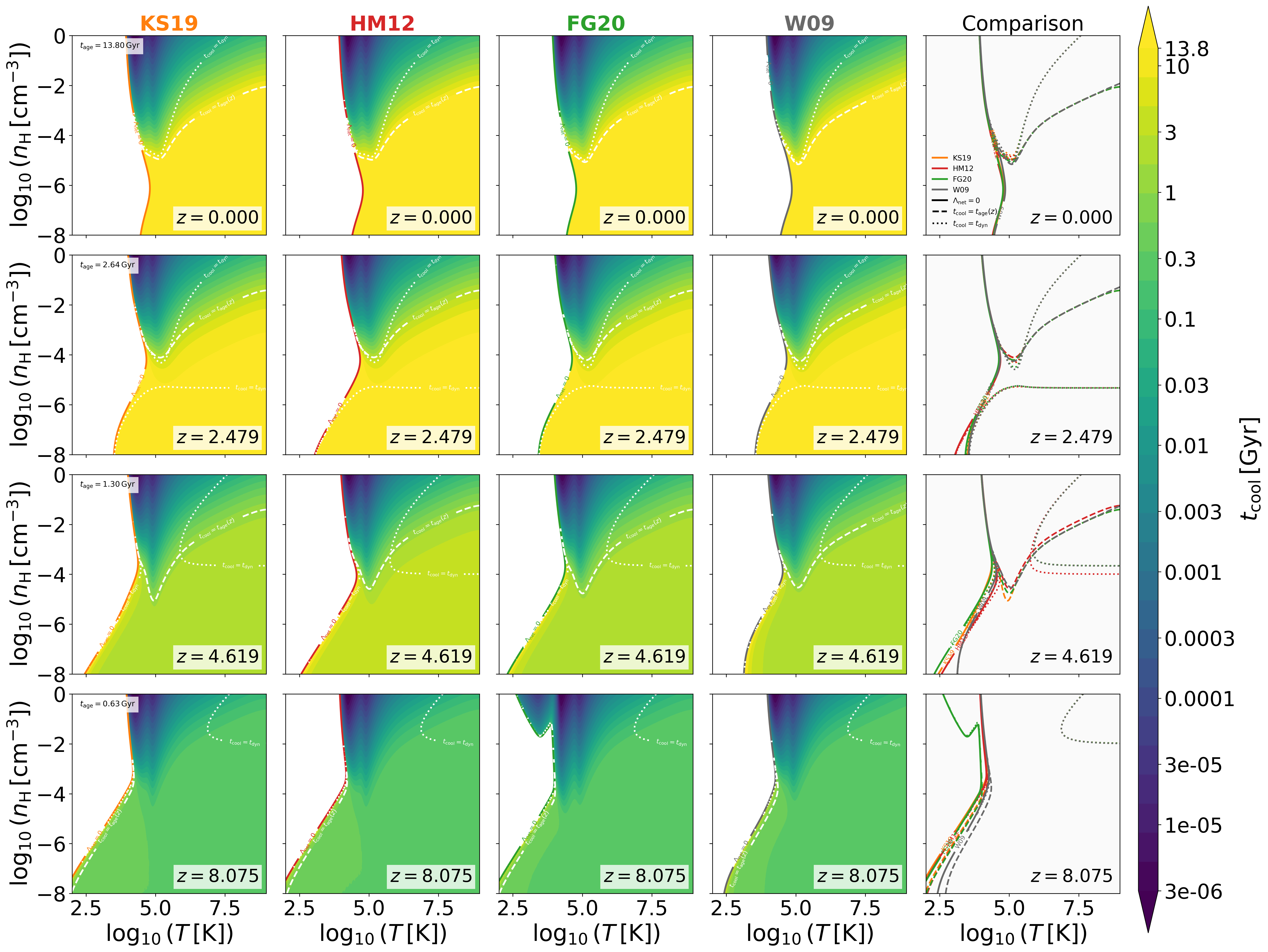}
    \caption{
    Same as Figure ~\ref{fig:tcool_solar}, but for H and He gas with
    $Y=0.248$. The persistence of model-dependent differences in the
    absence of metals shows that they do not arise exclusively from
    metal-line cooling.
    }
    \label{fig:tcool_hhe}
\end{figure*}

The H and He results in Figure ~\ref{fig:tcool_hhe} exhibit the same qualitative redshift evolution as the Solar-composition calculations. The UVB dependence is therefore already present through hydrogen and helium photoionization and photoheating. The differences are generally less extended in the density-temperature plane than in the Solar case, illustrating the additional role of metal-line cooling in shaping the thermal response.

\begin{figure*}
    \centering
    \includegraphics[width=0.82\textwidth]
    {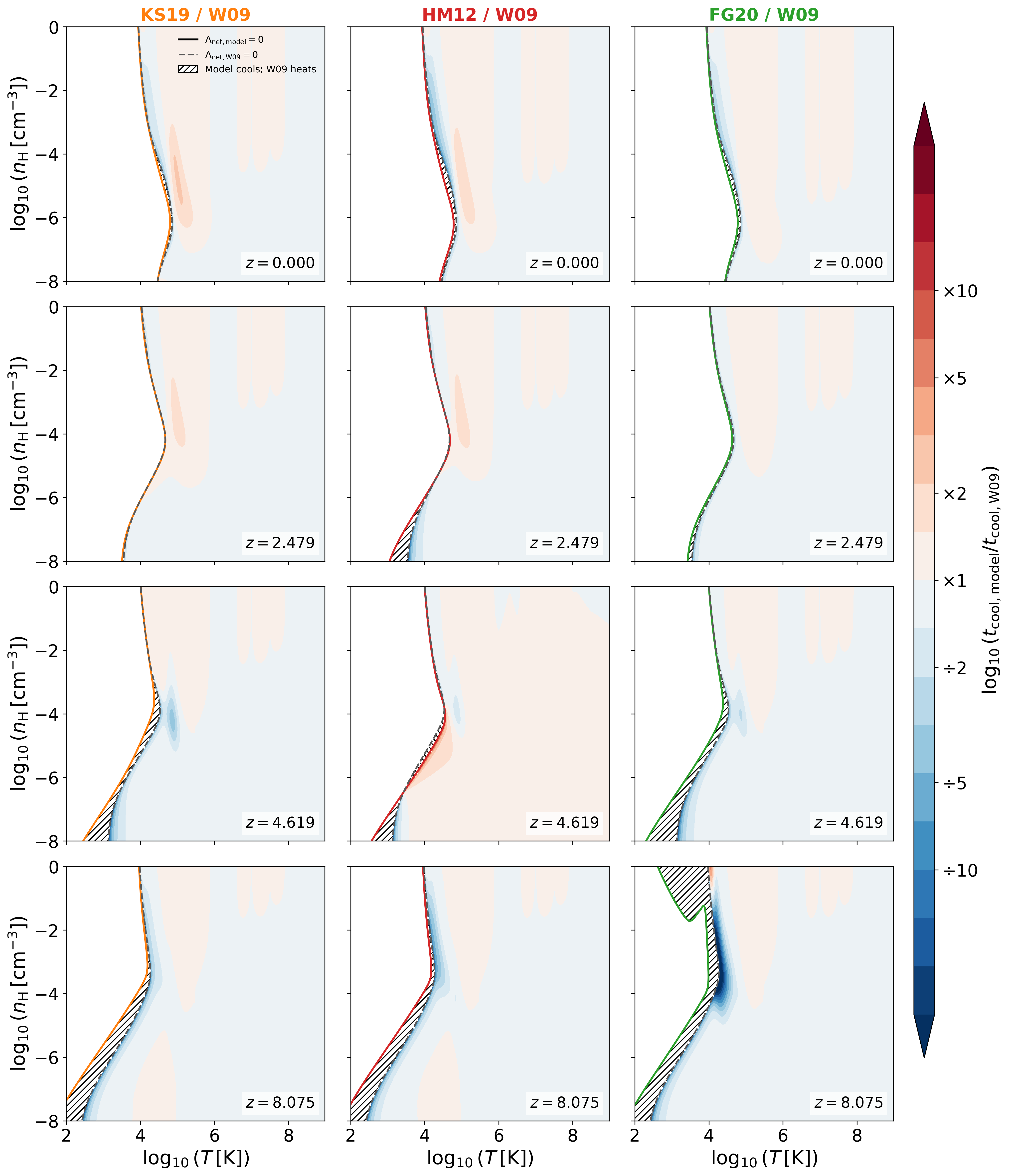}
    \caption{
    Same as Figure ~\ref{fig:tcool_ratio_solar}, but for H and He gas with
    $Y=0.248$.
    }
    \label{fig:tcool_ratio_hhe}
\end{figure*}

Figure~\ref{fig:tcool_ratio_hhe} shows the corresponding cooling-time differences relative to \wie. As for the Solar-composition analysis, this
comparison quantifies the combined impact of updating the table calculations, including changes in the radiation field, numerical implementation, and abundance set. Comparisons among KS19, HM12, and FG20 separately show the dependence on the adopted UVB within the new calculations.

Taken together, the hydrogen and helium equilibrium and cooling-time results are consistent with the principal conclusions obtained for the fixed Solar reference composition. Removing metals changes the magnitude and detailed extent of individual features, but it does not remove the underlying sensitivity of the thermal state to the adopted UVB prescription.

%\printcredits

\bibliographystyle{cas-model2-names}
\bibliography{Paper_UVB}

\end{document}